\documentclass[sigconf,nonacm]{acmart}

\usepackage{tikz}
\usepackage{amsmath}
\usepackage{todonotes}
\usepackage{xspace}
\usepackage[nolist]{acronym}
\usepackage{amsmath}
\usepackage{listings}
\usepackage{pifont}
\usepackage{import}
\usepackage[capitalise,noabbrev]{cleveref}
\usepackage{booktabs}
\usepackage{multirow}
\usepackage{balance}
\usepackage{url}
\usepackage{subcaption}
\usepackage{lstautogobble}
\usepackage{makecell}

\definecolor{dgreen}{rgb}{0.0, 0.5, 0.0}
\definecolor{rs-base03}{RGB}{28, 28, 28}
\definecolor{rs-base02}{RGB}{38, 38, 38}
\definecolor{rs-base01}{RGB}{78, 78, 78}
\definecolor{rs-base00}{RGB}{88, 88, 88}
\definecolor{rs-base0}{RGB}{128, 128, 128}
\definecolor{rs-base1}{RGB}{138, 138, 138}
\definecolor{rs-base2}{RGB}{215, 215, 175}
\definecolor{rs-base3}{RGB}{255, 255, 215}
\definecolor{rs-yellow}{RGB}{175, 135, 0}
\definecolor{rs-orange}{RGB}{215, 95, 0}
\definecolor{rs-red}{RGB}{215, 0, 0}
\definecolor{rs-magenta}{RGB}{175, 0, 95}
\definecolor{rs-violet}{RGB}{95, 95, 175}
\definecolor{rs-blue}{RGB}{0, 135, 255}
\definecolor{rs-cyan}{RGB}{0, 175, 175}
\definecolor{rs-green}{RGB}{95, 135, 0}
\lstdefinelanguage{Rust}
                   {morekeywords={as,break,const,continue,crate,else,enum,extern,false,
                       fn,for,if,impl,in,let,loop,match,mod,move,mut,pub,ref,return,Self,
                       self,static,struct,super,trait,true,type,unsafe,use,where,while,
                       abstract,alignof,become,box,do,final,macro,offsetof,override,priv,
                       proc,pure,sizeof,typeof,unsized,virtual,yield},
                     morekeywords=[2]{isize,usize,char,bool,str,String,u8,u16,u32,u64,u128,i8,i16,i32,i64,i128,f32,f64},
                     sensitive=true,
                     morecomment=[l]{//},
                     morecomment=[l]{///}, % change color
                     morestring=[b]{"},
                   }%

\lstdefinestyle{rust-code}{%
  language=Rust,                   % the language
  basicstyle=\footnotesize\ttfamily,   % size of the fonts for the code
  keywordstyle=\color{rs-yellow},       % core keywords
  keywordstyle={[2]\color{rs-blue}}, % built-ins
  stringstyle=\color{rs-red},
  commentstyle=\color{rs-green},
  upquote=true,                      % requires textcomp
  frame=single,
}
\lstdefinestyle{CStyle}{language=C++,
                basicstyle=\footnotesize\ttfamily,
                keywordstyle=\color{blue}\ttfamily,
                stringstyle=\color{red}\ttfamily,
                commentstyle=\color{green}\ttfamily,
                morecomment=[l][\color{magenta}]{\#},
                frame=single,
                }
\lstdefinestyle{XMLStyle}{morestring=[b]",
  morestring=[s]{>}{<},
  morecomment=[s]{<?}{?>},
  stringstyle=\color{black},
  identifierstyle=\color{darkblue},
  keywordstyle=\color{cyan},
  morekeywords={xmlns,version,type}% list your attributes here
                }

\newcommand{\pA}{\ding{182}\xspace}
\newcommand{\pB}{\ding{183}\xspace}
\newcommand{\pC}{\ding{184}\xspace}
\newcommand{\pD}{\ding{185}\xspace}
\newcommand{\pE}{\ding{186}\xspace}

\newcommand{\toolname}{\textsc{MetaEmu}\xspace}
\renewcommand{\textapprox}{\textasciitilde}

\setuptodonotes{inline}

\setcopyright{none} % No copyright notice required for submissions
\acmConference{Preprint}
\begin{document}

\begin{acronym}[ASCII]
 \acro{3DES}{Triple DES}
 \acro{AES}{Advanced Encryption Standard}
 \acro{ANF}{Algebraic Normal Form}
 \acro{API}{Application Programming Interface}
 \acro{ARX}{Addition, Rotate, XOR}
 \acro{ASK}{Amplitude-Shift Keying}
 \acro{ASIC}{Application Specific Integrated Circuit}
 \acro{BL}{Bootloader Enable}
 \acro{BOR}{Brown-Out Reset}
 \acro{BPSK}{Binary Phase Shift Keying}
 \acro{CBC}{Cipher Block Chaining}
 \acro{CBS}{Critical Bootloader Section}
 \acro{CGM}{Continuous Glucose Monitoring System}
 \acro{CMOS}{Complementary Metal Oxide Semiconductor}
 \acro{COPACOBANA}{Cost-Optimized Parallel Code Breaker and Analyzer}
 \acro{CPA}{Correlation Power Analysis}
 \acroplural{CPA}[CPAs]{Correlation Power Analyzes}
 \acro{CPU}{Central Processing Unit}
 \acro{CTR}{Counter \acroextra{(mode of operation)}}
 \acro{CRC}{Cyclic Redundancy Check}
 \acro{CRP}{Code Readout Protection}
 \acro{DES}{Data Encryption Standard}
 \acro{DC}{Direct Current}
 \acro{DDS}{Digital Direct Synthesis}
 \acro{DFT}{Discrete Fourier Transform}
 \acro{DMA}{Direct Memory Access}
 \acro{DoS}{Denial-of-Service}
 \acro{DFA}{Differential Fault Analysis}
 \acro{DPA}{Differential Power Analysis}
 \acro{DRAM}{Dynamic Random-Access Memory}
 \acro{DRM}{Digital Rights Management}
 \acro{DSO}{Digital Storage Oscilloscope}
 \acro{DSP}{Digital Signal Processing}
 \acro{DST}{Digital Signature Transponder}
 \acro{DUT}{Device Under Test}
 \acroplural{DUT}[DUTs]{Devices Under Test}
 \acro{ECB}{Electronic Code Book}
 \acro{ECC}{Elliptic Curve Cryptography}
 \acro{ECU}{Electronic Control Unit}
 \acro{EDE}{Encrypt-Decrypt-Encrypt \acroextra{(mode of operation)}}
 \acro{EEPROM}{Electrically Erasable Programmable Read-Only Memory}
 \acro{EM}{Electro-Magnetic}
 \acro{FFT}{Fast Fourier Transform}
 \acro{FCC}{Federal Communications Commission}
 \acro{FIR}{Finite Impulse Response}
 \acro{FIVR}{Fully Integrated Voltage Regulator}
 \acro{FPGA}{Field Programmable Gate Array}
 \acro{FSK}{Frequency Shift Keying}
 \acro{GIAnT}{Generic Implementation ANalysis Toolkit}
 \acro{GMSK}{Gaussian Minimum Shift Keying}
 \acro{GPIO}{General Purpose I/O}
 \acro{GPR}{General Purpose Register}
 \acro{HD}{Hamming Distance}
 \acro{HDL}{Hardware Description Language}
 \acro{HF}{High Frequency}
 \acro{HAL}{Hardware Abstraction Layer}
 \acro{HMAC}{Hash-based Message Authentication Code}
 \acro{HW}{Hamming Weight}
 \acro{IC}{Instrument Cluster}
 \acro{ID}{Identifier}
 \acro{ISM}{Industrial, Scientific, and Medical \acroextra{(frequencies)}}
 \acro{IIR}{Infinite Impulse Response}
 \acro{IL}{Intermediate Language}
 \acro{IP}{Intellectual Property}
 \acro{IoT}{Internet of Things}
 \acro{IR}{Intermediate Representation}
 \acro{ISR}{Interrupt Service Routine}
 \acro{IV}{Initialization Vector}
 \acro{JTAG}{Joint Test Action Group}
 \acro{LF}{Low Frequency}
 \acro{LFSR}{Linear Feedback Shift Register}
 \acro{LQI}{Link Quality Indicator}
 \acro{LSB}{Least Significant Bit}
 \acro{LSByte}{Least Significant Byte}
 \acro{LUT}{Look-Up Table}
 \acro{MAC}{Message Authentication Code}
 \acro{MCU}{Microcontroller Unit}
 \acro{MF}{Medium Frequency}
 \acro{MMU}{Memory Management Unit}
 \acro{MITM}{Man-In-The-Middle}
 \acro{MSB}{Most Significant Bit}
 \acro{MSByte}{Most Significant Byte}
 \acro{MSK}{Minimum Shift Keying}
 \acro{MSR}{Model Specific Register}
 \acro{muC}{\mu Microcontroller}
 \acro{NLFSR}{Non-Linear Feedback Shift Register}
 \acro{NLF}{Non-Linear Function}
 \acro{NFC}{Near Field Communication}
 \acro{NRZ}{Non-Return-to-Zero \acroextra{(encoding)}}
 \acro{NVM}{Non-Volatile Memory}
 \acro{OOK}{On-Off-Keying}
 \acro{OP}{Operational Amplifier}
 \acro{OTP}{One-Time Password}
 \acro{PC}{Personal Computer}
 \acro{PCB}{Printed Circuit Board}
 \acro{PhD}{Patiently hoping for a Degree}
 \acro{PKE}{Passive Keyless Entry}
 \acro{PKES}{Passive Keyless Entry and Start}
 \acro{PKI}{Public Key Infrastructure}
 \acro{PMBus}{Power Management Bus}
 \acro{PoC}{Proof-of-Concept}
 \acro{POR}{Power-On Reset}
 \acro{PPC}{Pulse Pause Coding}
 \acro{PRNG}{Pseudo-Random Number Generator}
 \acro{PSK}{Phase Shift Keying}
 \acro{PWM}{Pulse Width Modulation}
 \acro{RAPL}{Running Average Power Limit}
 \acro{RDP}{Read-out Protection}
 \acro{RF}{Radio Frequency}
 \acro{RFID}{Radio Frequency IDentification}
 \acro{RKE}{Remote Keyless Entry}
 \acro{RNG}{Random Number Generator}
 \acro{ROM}{Read Only Memory}
 \acro{ROP}{Return-Oriented Programming}
 \acro{RSA}{Rivest Shamir and Adleman}
 \acro{RTL}{Register Transfer Language}
 \acro{SCA}{Side-Channel Analysis}
 \acro{SDR}{Software-Defined Radio}
 \acro{SGX}{Software Guard Extensions}
 \acro{SNR}{Signal to Noise Ratio}
 \acro{SHA}{Secure Hash Algorithm}
 \acro{SHA-1}{Secure Hash Algorithm 1}
 \acro{SHA-256}{Secure Hash Algorithm 2 (256-bit version)}
 \acro{SMA}{SubMiniature version A \acroextra{(connector)}}
 \acro{SMBus}{System Management Bus}
 \acro{I2C}{Inter-Integrated Circuit}
 \acro{ISA}{Instruction Set Architecture}
 \acro{SPA}{Simple Power Analysis}
 \acro{SMT}{Satisfiability Modulo Theories}
 \acro{SPI}{Serial Peripheral Interface}
 \acro{SPOF}{Single Point of Failure}
 \acro{SoC}{System on Chip}
 \acro{SVID}{Serial Voltage Identification}
 \acro{SWD}{Serial Wire Debug}
 \acro{TCB}{Trusted Computing Base}
 \acro{TCU}{Telematics Control Unit}
 \acro{TEE}{Trusted Execution Environment}
 \acro{TMTO}{Time-Memory Tradeoff}
 \acro{TMDTO}{Time-Memory-Data Tradeoff}
 \acro{TZ}[TrustZone]{TrustZone}
 \acro{RSSI}{Received Signal Strength Indicator}
 \acro{SHF}{Superhigh Frequency}
 \acro{UART}{Universal Asynchronous Receiver Transmitter}
 \acro{UCODE}{\mu Microcode}
 \acro{UHF}{Ultra High Frequency}
 \acro{UID}{Unique Identifier}
 \acro{USRP}{Universal Software Radio Peripheral}
 \acro{USRP2}{Universal Software Radio Peripheral (version 2)}
 \acro{USB}{Universal Serial Bus}
 \acro{VHF}{Very High Frequency}
 \acro{VLF}{Very Low Frequency}
 \acro{VHDL}{VHSIC (Very High Speed Integrated Circuit) Hardware Description Language}
 \acro{VR}{Voltage Regulator}
 \acro{VXE}{Virtual Execution Environment}
 \acro{WLAN}{Wireless Local Area Network}
 \acro{XOR}{Exclusive OR}
 \acro{IR}{Intermediate Representation}
 \acro{OCD}{On-Chip Debug}
 \acro{BCM}{Body Control Module}
 \acro{GCM}{Gateway Control Module}
 \acro{CAN}{Controller Area Network}
 \acro{MMIO}{Memory-mapped I/O}
 \acro{UDS}{Unified Diagnostic Services}
 \acro{CMT}{Compare and Match Timer}
 \acro{GPT}{General Purpose Timer}
\end{acronym}

\date{}

\title{\toolname: An Architecture Agnostic Rehosting Framework for Automotive Firmware}

\author{Zitai Chen}
\authornote{Both authors contributed equally to this research.}
\email{z.chen@pgr.bham.ac.uk}
\affiliation{%
  \institution{University of Birmingham}
  \country{United Kingdom}
  \postcode{B15 2TT}
}
\author{Sam L. Thomas}
\authornotemark[1]
\authornote{Research partially undertaken while at University of Birmingham.}
\email{sam@binarly.io}
\affiliation{%
  \institution{BINARLY, Inc.}
  \country{United States of America}
}
\author{Flavio D. Garcia}
\email{f.garcia@cs.bham.ac.uk}
\affiliation{%
  \institution{University of Birmingham}
  \country{United Kingdom}
  \postcode{B15 2TT}
}

\begin{abstract}
In this paper we present \toolname, an architecture-agnostic emulator synthesizer geared towards rehosting and security analysis of automotive firmware.
\toolname improves over existing rehosting environments in two ways:
Firstly, it solves the hitherto open-problem of a lack of generic \acp{VXE} for rehosting by synthesizing processor simulators from Ghidra's language definitions. In doing so, \toolname can simulate any processor supported by a vast and growing library of open-source definitions. In \toolname, we use a specification-based approach to cover peripherals, execution models, and analyses, which allows our framework to be easily extended.
Secondly, \toolname can rehost and analyze multiple targets, each of different architecture, simultaneously, and share analysis facts between each target's analysis environment, a technique we call inter-device analysis.

We show that the flexibility afforded by our approach does not lead to a performance trade-off---\toolname lifts rehosted firmware to an optimized intermediate representation, and provides performance comparable to existing emulation tools, such as Unicorn. Our evaluation spans five different architectures, bare-metal and RTOS-based firmware, and three kinds of automotive \ac{ECU} from four distinct vendors---none of which can be rehosted or emulated by current tools, due to lack of processor support. Further, we show how \toolname enables a diverse set of analyses by implementing a fuzzer, a symbolic executor for solving peripheral access checks, a CAN ID reverse engineering tool, and an inter-device coverage tracker.

\end{abstract}

\ccsdesc[300]{Hardware~Post-manufacture validation and debug}
\ccsdesc[300]{Hardware~Simulation and emulation}
\ccsdesc[500]{Software and its engineering~Software reverse engineering}
\ccsdesc[500]{Software and its engineering~Dynamic analysis}
\ccsdesc[500]{Computer systems organization~Firmware}
\ccsdesc[400]{Security and privacy~Embedded systems security}
\ccsdesc[500]{Computer systems organization~Embedded software}

\keywords{automotive, dynamic program analysis, firmware, emulation}

\maketitle

\section{Introduction}

Automobiles fulfill a vital function in our societies---a means of transport---and, as with many technologies, have become increasingly connected, and at the same time more closed to scrutiny~\cite{dieselgate}. Kocher et al.~\cite{conf/sp/KoscherCRPKCMKASS10} and Checkoway et al.~\cite{conf/uss/CheckowayMKASSKCRK11} demonstrated that this increased connectivity introduces a significant attack surface,
and, inspired by these seminal works, Miller and Valasek~\cite{remotejeep} showed that automobiles, like any other Internet-connected device can be remotely compromised. Meanwhile, Garcia et al.~\cite{conf/uss/GarciaOKP16} and Verdult et al.~\cite{conf/uss/VerdultGB12,conf/uss/VerdultGE13} showed that physical entry to a vehicle---something traditionally handled by a mechanical means---could also be defeated.

While researchers have sought to mitigate these flaws by protocol enhancements, e.g.,~\cite{conf/acsac/BulckMP17}, analyzing the black-box firmware of the \acp{ECU} connected to a vehicle's internal network, has largely been left as a manual endeavor, e.g.,~\cite{conf/esorics/HerrewegenG18}. We posit that this is due to the fact that, while \ac{ECU} firmware is not necessarily hard to obtain, tool support for performing anything more than manual analysis is severely lacking for the processor architectures they are based on.

Recent approaches enabling security analysis of embedded devices have primarily focused on rehosting, i.e., the process of transplanting a device's firmware to run inside a virtualized execution environment, e.g., QEMU~\cite{qemu} or derived tools, such as Unicorn~\cite{unicorn}, to enable specific analysis tasks. While these environments support a wide-range of commonly used embedded architectures, such as ARM and MIPS, they do not provide out-of-the-box support for arbitrary peripherals, or any support for esoteric architectures, such as those found in automotive components. At the time of writing, most published work has sought to address the former challenge: peripheral support. However, as noted by Fasano et al.~\cite{conf/asiaccs/FasanoBMLBDEFLG21} in their systematization of the field, for devices whose firmware is not supported by an off-the-shelf emulator, the latter challenge---obtaining a suitable execution environment---remains an open problem, hampering the analysis of a large and vital class of devices.

An orthogonal issue arising from the use of commodity emulators for rehosting is that they implicitly force a device-centric approach, where a device and its peripherals are assumed to operate with little or no constraints on the inputs they receive. For many scenarios where embedded devices are deployed, we believe that this simplification is unjustified, as evidenced in recent work by industry practitioners~\cite{keenlab1,keenlab2}. In the automotive setting, for example, ECUs are interconnected by a \ac{CAN} bus, which they use to communicate. Many ECUs require the presence of other ECUs to successfully initialize and operate, hence, to realistically simulate a device's operating environment, we often require more than a single ECU to be rehosted.

\textbf{Our contribution} In this paper, we address the challenge of analyzing devices that cannot be rehosted using commodity emulators. We present our framework, \toolname, that takes as input widely available processor and instruction set definitions---Ghidra's language definitions, and automatically synthesizes virtualized execution environments capable of rehosting multiple devices of different architectures simultaneously. Our synthesized environments enable deep introspection of each rehosted firmware's state and facilitate complex inter-device analyses.

We demonstrate that our approach is general (it can support rehosting firmware of many esoteric architectures, irrespective of their peculiarities), scalable (it can analyze many instances of a firmware in parallel and share analysis facts), and enables dynamic analysis of a class of devices that until now have been largely overlooked.

To evaluate \toolname, we benchmark its performance and provide six case studies detailing its use for different kinds of security analysis, including backdoor detection, inter-device analysis, and various automotive-focused reverse engineering tasks. For these analyses, we implement a fuzzer (by integrating \toolname with LibAFL~\cite{libafl}), a symbolic executor, which we use to solve peripheral access checks, a \ac{CAN} bus reverse-engineering tool, and an inter-firmware coverage tracker. Our case studies demonstrate that \toolname is effective in analyzing complex, binary-only firmware, without any architecture support or performance trade-off. Our evaluation data-set consists of thirteen benchmarks and six device firmware: two based on open-source SDKs and four extracted from automotive \acp{ECU}: a \ac{BCM}, an \ac{IC}, and two \acp{TCU}. Each end-user firmware is based on a different CPU architecture (Infineon C166 and Renesas RH850, SH-2A, and V850E2M-M), and each presents a different set of challenges for emulation and analysis.

To summarize, our work contributes to the state-of-the-art in the following ways:

\begin{enumerate}
    \item We present, to the best of our knowledge, the first generic framework for rehosting and dynamically analyzing end-user automotive firmware, and the first framework supporting device inter-dependent analysis of multiple rehosted firmware executing simultaneously.
    \item We show that our framework is capable of synthesizing \acp{VXE} for multiple esoteric architectures, generically enabling analysis of a vital class of devices currently unsupported by any other dynamic analysis framework.
    \item We show how our framework enables complex analyses, such as symbolic execution and fuzz testing, to be applied to binary-only automotive firmware.
    \item We provide an in-depth evaluation of \toolname, with respect to performance, implementation flexibility, and real-world usability. It spans five different architectures, bare-metal and RTOS-based firmware, and three kinds of automotive \ac{ECU} from four distinct vendors. %and four end-user firmware wherein we rehost and perform dynamic analysis of four end-user \ac{ECU} firmware.
\end{enumerate}

We release our framework, processor definitions, and firmware data-set as open-source---for details see~\cref{sec:open-source}. %We provide an example of how to use \toolname in \cref{sec:can_id}.

\section{Background}
\label{sec:background}

In this section, we provide the necessary background to understand our contributions. We first provide an overview of the challenges involved in analyzing automotive components, and then cover the terminology used in the rest of this article.

A modern vehicle is composed of multiple \acp{ECU} connected by a \ac{CAN} bus. Each \ac{ECU} may have multiple peripherals, and often, each will be manufactured by a different vendor, and may be based on a different CPU architecture. In contrast to the majority of devices analyzed under the umbrella of ``embedded device security'' in the literature, \acp{ECU} are generally based on more esoteric architectures, e.g., V850/RH850, or C166, rather than ARM or MIPS. Unfortunately, these architectures are not supported at all by the de facto emulation environments for security analysis---QEMU and Unicorn. This makes it impossible to apply modern security analysis approaches such as symbolic execution to these devices. While one might assume that this challenge can be addressed by adding support for new architectures to those tools, doing so requires an unreasonably large amount of engineering effort: 1000s of lines of code and many months of development time---an unjustifiable amount of work for most applications. While Ghidra has \texttt{EmulatorHelper} which can be used to emulate some of these architectures, it does not have an easy-to-use interface for peripheral handling or adequate performance for intensive analyses.

\begin{figure*}[t!]
\centering
\resizebox{0.7\textwidth}{!}{
\def\svgwidth{0.7\textwidth}
\footnotesize
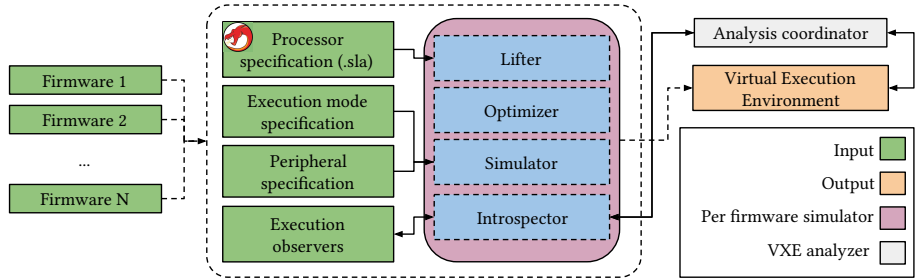
}
\caption{Overview of \toolname. Each firmware along with corresponding processor, execution mode, and peripheral specifications are used to synthesize a simulator. Each simulator (pink box) can have execution observers attached, which monitor and manipulate its execution state and interface with an analysis coordinator (gray box) to facilitate inter-firmware interaction and analyses. The resulting \ac{VXE} (yellow box) is composed of many per firmware simulators operating in parallel. \label{fig:overview}}

\vspace{-1em}
\end{figure*}

The tangential problem of peripheral support is a significant challenge in the automotive context: on the one hand, due to the variety and number of peripherals a single firmware might interface with, and on the other, due to a lack of documentation.  For some analysis tasks, peripheral interactions can be \emph{bypassed} by providing a satisfying value which will work even with limited information about the peripheral, as is done in previous work~\cite{conf/acsac/CaoGM020,conf/uss/0026G0Z21,conf/uss/FengML20}. However, such a bypass is often no better than (unsoundly) forcing execution of a given branch target. This is because the \emph{real} constraints on the peripheral register might depend on another device's output, or be constrained outside the execution path being analyzed.
An obvious example of such a peripheral is a \ac{CMT}, which is often used by firmware to implement task scheduling. Clearly, providing \emph{any} satisfying value for the peripheral's output will lead to undesired behavior, and is better handled by a modeling approach, such as that proposed by Gustafson et al.~\cite{conf/raid/GustafsonMSRMFB19}.
Thus, support for peripheral models and the ability to have multiple rehosted firmware interact and communicate is vital to ensure faithful simulation.

\toolname provides a means to rehost one or more device firm\-ware inside a virtual execution environment. We use the term rehost to jointly refer to the process of transplanting a device's firm\-ware into an emulator, and the simulation of its execution and interaction with its peripherals. In this work, we refer to the rehosting environment as a \ac{VXE}, and assume that a \ac{VXE} facilitates multiple firm\-ware to coexist and interact with each other. \toolname synthesizes \acp{VXE}; we distinguish a synthesized \ac{VXE} from a standard \ac{VXE}, such as those built on top of Unicorn~\cite{unicorn}, e.g., ~\cite{conf/uss/RugeCGH20,conf/wisec/MaierSP20}, by how they are specified: synthesized environments are specified using a declarative approach, rather than purely programmatically.

\section{System Overview}
\label{sec:overview}

In this section we provide an overview of \toolname, its inputs, and the assumptions we make about the firmware it can be used to rehost.

\subsection{Framework Architecture}\label{sec:framework_arch}

We summarize \toolname's architecture in Figure~\ref{fig:overview}. \toolname takes as input one or more binary blob firmware and outputs a combined simulation/analysis environment capable of rehosting all the input firmware within a single \ac{VXE}.
In the combined \ac{VXE}, each firmware has its own simulator (pink box in Figure~\ref{fig:overview}). To synthesize these simulators, \toolname requires four inputs: \pA a processor specification, \pB an execution mode specification, \pC a list of execution observers, and \pD a peripheral specification. To dynamically introspect and manipulate the firmware's state, \toolname attaches user-defined \emph{observers} to the simulator, which can be triggered on, e.g., register reads and memory writes. The attached observers enable us to handle complex addressing modes (e.g., that used for the C166 architecture in \S\ref{sec:c166}) and implement symbolic solvers for peripheral checks (similar to $\mu Emu$~\cite{conf/uss/0026G0Z21} and Laelaps~\cite{conf/acsac/CaoGM020}). By interfacing with the analysis coordinator (gray box), observers can communicate with other rehosted devices (e.g., allowing one device to provide input to another's peripherals) and share analysis facts, enabling inter-device analyses.

\subsubsection*{Framework Inputs}

\toolname is capable of rehosting many different types of devices; from those with bare-metal firmware to those based on embedded RTOSes, such as ThreadX. Regardless of the type of firmware being rehosted, we make the following assumptions:

\begin{enumerate}
\item We do not have access to the firmware's source-code.
\item We know its basic memory map and architecture.
\item We know the memory regions it uses for memory-mapped I/O with its peripherals.
\item We have a Ghidra language definition for its architecture.
\end{enumerate}

Our assumptions are con\-sis\-tent with exist\-ing approaches, such as P$^2$IM~\cite{p2imdocs,conf/uss/FengML20} and $\mu$Emu~\cite{uemudocs, conf/uss/0026G0Z21}, which also require manual specification of the memory mappings for each target firmware/device.

In addition to the firmware to rehost, \toolname requires four auxiliary inputs to synthesize a \ac{VXE} (shown in green in \cref{fig:overview}).

\smallskip
\noindent\textbf{Processor specification:} To disassemble and obtain an architecture independent representation of each firm\-ware, \toolname leverages Ghidra's language definitions. These consist of a \emph{pspec} which defines register name and basic memory mappings, a \emph{cspec} which defines calling conventions and size information of primitive types, and a \emph{slaspec} which describes how to disassemble the architecture's instructions and an encoding of their semantics. At the time of writing, there are 28 processor families supported in Ghidra master branch and many others created by the community.

\smallskip
\noindent\textbf{Execution mode specification:} \toolname can simulate each firm\-ware's execution using different strategies (further detailed in \S\ref{sec:execution-modes}), including concrete execution and concolic execution. Each specification defines how \toolname simulates firmware at the level of individual \ac{IL} operations and how we represent the firmware's state. All mode specifications implement a common interface and each synthesized firmware simulator is parameterized by a generic type constrained by that interface, allowing it to be instantiated to operate under any execution mode.

\smallskip
\noindent\textbf{Execution observers:} To extend each synthesized simulator to support architectural nuances, peripherals, interrupts, and user-defined analyses, we use so-called observers. As discussed later in \S\ref{sec:observers}, our observers intercept events during simulation specific to the simulator's execution mode, and in response, can modify the firmware's state and coordinate with other simulators and observers through a \ac{VXE}-shared \emph{analysis coordinator}.

\smallskip
\noindent\textbf{Peripheral specification:} As detailed later in \S\ref{sec:peripherals}, \toolname supports peripherals through generic and user-specified models (\S\ref{sec:universal-peripheral-backend}), and automatic model inference (\S\ref{sec:peripheral-solving}). We specify all types of peripheral using execution observers, which we attach to simulators via their introspection interface. This interface is flexible and can support peripherals that receive input from outside of a \ac{VXE}, enabling us to interact with rehosted firmware via SocketCAN, fuzzers such as AFL~\cite{afl}, and even physical devices.

\subsection{Challenges Rehosting Complex Firmware}

Fundamentally \toolname is a multi-target emulation framework purpose-built to facilitate complex security analysis that can be customized and extended using a specification-based approach to alter execution policies, architecture definitions, and peripheral models. We encountered several challenges during the design and implementation of \toolname. In this section, we provide an overview of these difficulties and how we have addressed them.

\textbf{Architecture support and IR optimization:}
To facilitate architecture agnostic analysis, \toolname emulates a firmware by first translating its architecture-specific instructions to an intermediate representation, and performs emulation of the resulting \ac{IR} operations. The emitted operations are based on Ghidra's SLEIGH language definitions for the target's architecture. These definitions produce an IR representation that is not well suited for emulation, as it is overly verbose, as shown in Figure~\ref{fig:bad-il}. This is particularly problematic when using a symbolic execution policy, as a more complex IR leads to more complex symbolic formula.
To improve the performance, we implement a \ac{IR} optimizer in our lifter, which can produces a much more optimal \ac{IR} representation, as shown in Figure~\ref{fig:good-il}.

While our lifting and optimization approach is mostly standard, i.e., optimizations are applied on the SSA form representation of our IR, due to our use of SLEIGH specifications to generate our IR, we encountered a number of additional challenges in our implementation. The first being that Ghidra's language definitions allow for variables (both registers and temporaries) to overlap. SLEIGH represents each variable as a $(offset, size)$ pair that indexes into a flat address space, e.g., the register \texttt{RAX} might have offset $0$ and size $8$, while \texttt{EAX} might have offset $4$ and size $4$. This complicates SSA transformation, as it assumes that variables do not overlap. We thus, modify the standard algorithm to treat overlapping variables as equivalent. A further challenge is that SLEIGH's temporaries have a liveness that only spans a single instruction's IL operations. Hence, the same temporaries may be reused between consecutive instructions. We adapt the standard liveness data-flow analysis to account for this implicit lifetime information by providing unique labels for temporaries across instructions, thus preventing data-flow information propagating over instruction boundaries.

\textbf{Missing analysis state and context:}
When emulating embedded firmware, we often have to deal with missing or inconsistent state. For example, the values of a device's peripheral status registers, or global variables that may be have been set to particular values when  the firmware was dumped, which prevent the firmware executing correctly when starting emulation from another location. Under these circumstances, the emulated firmware will usually stall in a check-and-wait loop. To solve this problem automatically, we implement a peripheral check solver, which \pA identifies \emph{looping} states and \pB updates the firmware's context with suitable values to allow execution to proceed. The main difficulty is how to detect such states during emulation. A naive approach would be to track and count the number of times blocks of operations are executed and if a particular block is executed repeatedly over a threshold number of times, we assume that the emulated firmware has entered a stall-state. While such an approach works, in practice, it is both inefficient and requires a significant amount of resources. To efficiently address this in \toolname, we use a state machine-based approach, shown in Figure~\ref{fig:prprsolver} to detect stall-states based on both code and memory access patterns, which reduces the amount of program state to track simultaneously.

While our peripheral solver can be used to bypass checks that do not impact the behavior of the target firmware with respect to a particular analysis, when the data or functionality of the peripheral does matter for a given analysis, we cannot simply bypass a stall-state check with \emph{any} satisfying state configuration. Examples of this include the CAN peripheral of our RH850-based firmware (\S\ref{sec:rh850}) and the Compare and Match Timer of our C166-based firmware (\S\ref{sec:c166}). In such scenarios, the peripheral's behavior needs to be explicitly modeled. As even SoCs from the same family have slightly different implementations of such peripherals, and different kinds of context switching logic, it is both error-prone and time-consuming to implement versions of the same peripheral for different devices. To address this issue, we provide a universal peripheral backend (\S\ref{sec:universal-peripheral-backend}) and interrupt model. They abstract the peripheral behaviors and provide building blocks for implementing peripherals for different devices. Our interrupt framework enables a means to rapidly implement interrupt status tracking, context switching, and interrupt handler hooks and overrides.

\textbf{Inter-device analysis: } As mentioned in \S\ref{sec:background}, a micro-controller may rely on the data from other chips or devices to function properly. There are two sets of challenges faced when performing such inter-device analysis. Firstly, devices being rehosted in the same \ac{VXE} may come from different manufactures, have different methods of handling peripherals, and may be based on different architectures. Secondly, the devices may lack a documented shared channel for communication, that would otherwise be facilitated by \emph{black-box} hardware in a real-world setting. In \toolname, we overcome the first set of challenges by providing a wide-range of architecture support and three methods to implement peripheral models. We address the second challenge by providing an analysis coordinator that facilitates message passing using a common interface between both device's emulated peripherals.

\section{Implementation of \toolname}
\definecolor{lightGrey}{rgb}{0.9, 0.9, 0.9}
\newcommand{\Hilight}{\makebox[0pt][l]{\color{lightGrey}\rule[-0.3em]{\linewidth}{1.2em}}}

\begin{figure*}[ht!]
% Have to put it here to keep it in page 4
    \centering
    \footnotesize
    \begin{subfigure}{0.96\columnwidth}
    \begin{lstlisting}[numbers=left, frame=single, escapechar=\%, autogobble]
%\color{dgreen}\texttt{// mov dword [RBP - 0xc], ECX}%
var0620 := RBP + 0xfffffffffffffff4
%\Hilight%var1790 := ECX           %\color{dgreen}\texttt{// redundant variable var1790}%
%\Hilight%*var0620 := var1790
CF := 0x0
OF := 0x0
%\color{dgreen}\texttt{// xor ECX, ECX}%
%\Hilight%ECX := ECX ^ ECX         %\color{dgreen}\texttt{// ECX is constant value 0x0}%
%\Hilight%RCX := zext(ECX, 64)     %\color{dgreen}\texttt{// Zext of 0 is constant 0x0}%
%\Hilight%SF := ECX s< 0x0         %\color{dgreen}\texttt{// Constant 0x0}%
%\Hilight%ZF := ECX == 0x0         %\color{dgreen}\texttt{// Constant 0x1}%
%\Hilight%var2580 := ECX & 0xff    %\color{dgreen}\texttt{// Constant 0x0}%
%\Hilight%var2590 := popcount(var2580) %\color{dgreen}\texttt{//number of 1s in 0 is 0 }%
%\Hilight%var25a0 := var2590 & 0x1 %\color{dgreen}\texttt{// Constant 0x0}%
%\Hilight%PF := var25a0 == 0x0     %\color{dgreen}\texttt{// Constant 0x1}%
    \end{lstlisting}
    \caption{\ac{IR} produced by Ghidra's lifter. The highlighted lines indicate instructions that can be simplified.\label{fig:bad-il}}
    \end{subfigure}
    \hfill
    \begin{subfigure}{0.96\columnwidth}
    \begin{lstlisting}[numbers=left, frame=single, escapechar=\%, autogobble]
var0620 := RBP + 0xfffffffffffffff4
*var0620 := ECX
CF := 0x0
OF := 0x0
ECX := 0x0
RCX := 0x0
SF := 0x0
ZF := 0x1
PF := 0x1
    \end{lstlisting}
    \caption{Optimized \ac{IR} produced by \toolname for the sequence in Figure~\ref{fig:bad-il}. The optimization to ``\texttt{mov dword [RBP - 0xc], ECX}'' eliminates the intermediate variable \texttt{var1790}. The optimization to ``\texttt{xor ECX, ECX}'' uses an identity for XOR, i.e., $v \oplus v = 0$. Our optimizations are \emph{safe} and preserve both memory and register side-effects. \label{fig:good-il}}
    \end{subfigure}
    \vspace{-1em}
    \caption{Generated \ac{IR} for the x86-64 instructions: \texttt{mov dword [RBP - 0xc], ECX; xor ECX, ECX}.}
\vspace{-1em}
\end{figure*}

In this section, we describe \toolname's key features in depth and detail how they can be used to analyze rehosted firmware.
\toolname is written entirely in Rust and spans \textapprox 40kloc (excluding comments and dependencies). This includes \textapprox 24kloc for \emph{IR lifting and optimization}, \textapprox 5kloc for \emph{IR simulation and observers}, and \textapprox 11.5kloc for \emph{symbolic execution and other execution policy backends}. We do not rely on Ghidra's code-base for any functionality, other than its XML-based processor specifications.

\subsection{Lifting \& IR Generation \label{sec:impl-ir-lift}}
To support emulation in an architecture agnostic way, \toolname operates on an \ac{IR} that explicitly models all processor instructions and their side effects. We represent each architectural instruction as one or more \ac{IL} operations, which we obtain by a process called lifting, i.e., the translation of bytes into \ac{IL} operations. Our \ac{IR} is inspired by Ghidra's~\cite{ghidra} \ac{IR}, P-Code~\cite{pcode}.
In contrast to P-Code, our IR uses two isomorphic \ac{IL} encodings: a SSA-form expression-based variant, which we use for optimization, and a \ac{RTL} variant, semantically equivalent to P-Code, which we use for emulation.
Similar to P-Code, our IR supports arbitrary extensions via intrinsic operations, which we use to model dynamic processor state changes, such as localized address mode switches.

Our lifter takes as input unmodified Ghidra language definitions, which contain information about calling conventions and the necessary information to lift a byte stream into our \acp{IR}. This approach enables us to make use of the vast number of processor architecture definitions from Ghidra and greatly reduces the workload of adding support for new architectures to \toolname. Inspired by B2R2~\cite{conf/bar/jung2019}, our lifter is completely parallel, and supports lifting chunks of the same firmware across many cores.

While one might assume we could have used Ghidra's lifter as a basis for \toolname, unfortunately, it does not perform any optimization of the \ac{IR} it generates---it simply produces a literal translation of the specification for each instruction into \ac{IL} operations---which we find leads to as much as 25\% of the operations emitted being superfluous (\S\ref{sec:optmi-reduction}). Clearly, naively interpreting \ac{IL} operations emitted from Ghidra's lifter will be inefficient, and for analyses such as symbolic execution will lead to much more complex formula due to extraneous operations.
Figure~\ref{fig:bad-il} provides a taste of just how inefficient the generated \ac{IR} is.

\subsubsection*{IR Optimization} \label{sec:ir-optmize-impl}

In \toolname, we address these inefficiencies in two ways.
First, within our lifter, we rewrite and optimize our \ac{IR} prior to it being emitted and processed by our simulator. Second, we cache optimized IR blocks, both on disk and within a cache shared among all executor instances of the same firmware, which enables us to amortize the overhead of performing our optimizations. Accordingly, \toolname lifter can produce a much more optimal \ac{IR} representation, as shown in Figure~\ref{fig:good-il}.

We optimize \ac{IR} blocks by rewriting and simplification. To do so, we leverage e-graphs\footnote{An e-graph is a DAG representing IL terms combined with a congruence-closed equivalence relation over said terms.}~\cite{egraphPHD} to perform equality saturation~\cite{conf/popl/TateSTL09} on the SSA-form statements of each \ac{IR} block after applying rewrite rules. This enables us to obtain the most optimal representation of each statement with respect to
our rewrite rules.
For e-graph construction and manipulation, we use \texttt{egg}~\cite{2021-egg}.

We use our SSA-form expression-based \ac{IR}, to perform our optimizations, rather than our \ac{RTL}-based \ac{IR}, to ensure the correct application of our rewrite rules in the presence of instruction-local control-flow. Such control-flow occurs when the \ac{IR} corresponding to an architectural instruction requires loops or branches to model the architectural semantics fully (e.g., \texttt{rep stosb}). Thus, an \ac{IR} block is more akin to a sequence of instruction-level control-flow graphs, rather than a strict basic block.

After obtaining the SSA-form representation, we transform each statement into an e-graph, merge it into a block-level e-graph, and apply rewrite rules which cover algebraic identities and simplifications. Upon adding a new statement to the block-level e-graph, we greedily apply rewrite rules, and extract its most optimal representation with respect to AST depth (i.e., the shallower, the better). By extracting statements in this way, we preserve the original \ac{IR} ordering, and by constructing an e-graph for the entire block incrementally, we preserve the equivalence classes discovered, allowing each statement to benefit from any rewriting possibilities applied to its predecessors.
Finally, we use liveness analysis~\cite{dragonBook} to identify redundant assignments and remove them.

Our optimizations preserve all side-effects to memory and registers encoded in our unoptimized \ac{IR} and therefore will not negatively impact the accuracy of any downstream analyses performed. To the best of our knowledge, our use of e-graphs for optimizing IR for emulation is novel.
In evaluating our approach (\S\ref{sec:perf_bench}), we find that through IR optimization and designing our emulator specifically for rehosting, we achieve more than 400\% performance improvement over Ghidra's lifter and emulator.

\subsection{Firmware Simulation}\label{sec:firmw_simulation}

At the most fundamental level, the simulators \toolname generates load an input firmware, translate its instructions into our \ac{IR}, and interpret each emitted \ac{IL} operation. Within the output \ac{VXE}, each simulator runs in parallel on a separate thread, and communicates using a shared analysis coordinator.

\subsubsection{Execution Modes \& Policies}\label{sec:execution-modes}
The method each simulator uses for interpretation depends on the execution mode or policy it was synthesized with.
In contrast to traditional emulation environments, \toolname supports a variety of execution modes to simulate the execution of a rehosted firmware. Among other scenarios, these modes allow us to address the lack of complete peripheral models and support analyses bootstrapped using incomplete execution contexts, which often occur when analyzing firmware extracted from end-user devices.
Each policy also influences the kind of observations that attached execution observers (\S\ref{sec:observers}) can make and the underlying representation they act upon. We support five modes:

\smallskip
\noindent\pA \textbf{Concrete execution} follows the standard semantics of the firmware's \ac{ISA}. It represents state as a collection of byte arrays, and updates the state by directly interpreting each IL operation in a step-wise manner.

\smallskip
\noindent\pB \textbf{Concolic execution}~\cite{conf/kbse/Sen07a} permits us to mark some input variables, i.e., memory locations, as symbolic, while treating the remainder as in concrete execution. We obtain concrete assignments for symbolic variables by querying the \ac{SMT} solver, Boolector~\cite{DBLP:journals/jsat/NiemetzPB14}. By negating the path constraints associated with a branch, we can obtain an assignment that allows us to explore both branch targets. Under this policy observers can inject symbolic values into the firmware's state.

\smallskip
\noindent\pC \textbf{Forced execution}~\cite{conf/uss/PengDZXLS14} follows concrete execution semantics, with the caveat that at conditional branches, we can ``flip'' the branch condition, such that the non-satisfying branch target is taken.

\smallskip
\noindent\pD \textbf{Micro execution}~\cite{conf/icse/Godefroid14} follows concrete execution, with the exception our observers can intercept invalid memory accesses, map the locations into our \ac{VXE} state, and populate them by an analysis-dependent policy.

\smallskip
\noindent\pE \textbf{Flood execution}~\cite{conf/raid/WilhelmC07} follows forced execution, with the additional property that all branch targets, regardless of their satisfiability, are explored in accordance with an exploration policy, e.g., to a fixed bound or threshold branch coverage. In \toolname we implement a variant of flood execution based on micro execution, as proposed by Gotovchits et al.~\cite{conf/bar/gotovchits2018}. Under this policy, observers witness events from multiple execution paths simultaneously.

\subsubsection{Architectural Nuances}

To facilitate non-standard addressing, dynamic processor mode switches, and other architectural quirks, we provide an interface to implement \ac{IL} intrinsic operations. This interface enables \toolname to support functionality that cannot be directly modelled in our IR, e.g., instructions for cryptographic primitives, as well as modify how low-level operations such as memory reads and writes are performed. We use this interface to support the \emph{DPP override mechanism} address scheme for the C166 architecture~\cite{c166ref}, which requires us to alter how the simulator resolves memory accesses for a variable window of instructions.

\subsection{Observers \& Analysis Coordination}\label{sec:observers}

To extend our specification approach to support peripherals (\S\ref{sec:peripherals}), arbitrary extensions to our generated simulators (\S\ref{sec:interrupt-support} and \S\ref{sec:inte-extern-tools}), and different kinds of analyses (\S\ref{sec:case-studies}), we use a concept called \emph{execution observers}. Observers \emph{attach} to our simulators and register to receive event notifications via each simulator's \emph{introspector} (bottom blue box in Figure~\ref{fig:overview}). In response to events, we can configure observers to manipulate the running firmware's state and/or interface with the \ac{VXE} \emph{analysis coordinator} (gray box in Figure~\ref{fig:overview}) to enable inter-device communication or analyses.

As noted in \S\ref{sec:execution-modes}, the kind of events an execution observer can witness, and the firmware state representation it operates on, is tied to the \emph{execution policy} of the simulator it is attached to.
However, regardless of the execution policy a simulator is synthesized with, we can register observers to receive event notifications for a number of basic event kinds. These include: \textbf{register reads/writes}, \textbf{memory reads/writes}, \textbf{program counter changes}, \textbf{conditional branches}, and \textbf{function calls}. We associate each event kind with a corresponding response kind, which we use to facilitate common rehosting tasks, such as: instruction skipping, conditional branch flipping, function call replacement, and forking the simulator's execution state. We provide an exhaustive list of event kinds in Appendix~\ref{sec:observer_api}.

\subsubsection*{Inter-device analysis}
Our analysis coordinator provides a generic message passing interface that routes arbitrary event notifications between each simulator. It also acts as an abstract inter-device communication channel, that allows each rehosted device to communicate with any other device inside the same \ac{VXE}. For example, it enables us to easily connect devices by their \ac{HAL} APIs, without resorting to manually implementing low-level cross-device communication channels using emulated peripherals.

\subsection{Peripheral Support}
\label{sec:peripherals}
\begin{figure}[t]
  \centering
  \resizebox{0.7\columnwidth}{!}{
    \normalsize

    \def\svgwidth{0.6\textwidth}
    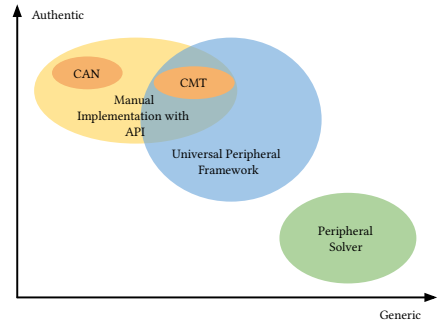

  }
    \vspace{-0.3cm}

  \caption{Types of peripheral support implemented in \toolname. \pA Peripheral Bypass (green) can be used to handle peripherals that only interact with firmware in a limited way---e.g., via status checks. This kind of peripheral handling does not faithfully model real-world peripheral behavior, but can be used to automatically bypass checks that would otherwise stall emulation. \pB Universal Peripheral models (blue) can be used to simulate \emph{fundamental} peripherals, e.g., timers, in hardware-agnostic manner, but require some manual configuration, hence are less general, but provide more true-to-hardware behavior. \pC Complete peripheral emulation (yellow) can be used to handle complex peripherals, such as those that perform I/O, e.g., over CAN, which require true-to-hardware behavior. These models cannot generalize beyond specific device families, and their behavior cannot be automatically inferred.
  \label{fig:peripheral_type}}
  \vspace{-2em}
\end{figure}

As discussed in \S\ref{sec:background}, correct and faithful firmware simulation depends on providing some degree of peripheral support. When rehosting, this support is still required: for example, to supply input to the firmware when fuzzing, or to ensure the firmware can execute at all if it is based on a RTOS.

In embedded firmware, peripheral interactions are frequently performed using \ac{MMIO}. To facilitate this, such firmware has a continuous memory region allocated specifically for \ac{MMIO}, which is divided into smaller regions for each peripheral. These regions contain memory mapped versions of a peripheral's control and status registers, as well as registers facilitating data transfers between the firmware and peripheral.
This presents two challenges when rehosting firmware: \pA When executing, firmware frequently checks its peripherals' statuses. If the status of a peripheral does not match what the firmware expects, e.g., due to being uninitialized, execution will ``stall'' in an infinite loop until the peripheral's status changes. Thus, without correct support, execution cannot proceed past such checks. \pB A firmware's control-flow is often dependent on data read from its peripherals. For example, for \acp{ECU} that communicate using \ac{UDS}, the specific service handler executed will be determined by data received via a \ac{CAN} peripheral. Hence, without peripheral support, it is impossible to faithfully explore such execution paths.

To overcome these obstacles, we provide three types of peripheral support in \toolname, as explained in \cref{fig:peripheral_type}. They are all based on execution observers (\S\ref{sec:observers}).
The first, inspired by Laelaps~\cite{conf/acsac/CaoGM020}, is a \textbf{generic Peripheral Check Solver} that allows us to specify a \ac{MMIO} address range and automatically bypass simple checking loops that would otherwise cause execution to stall. The second, is a \textbf{\emph{Universal} Peripheral framework}, which provides a specification-based approach to build common peripherals by composing generic models. These models are \textit{architecture-agnostic}, so can be reused for many different devices, with minimal reconfiguration. \toolname provides models to act as generic input sources, e.g., for fuzzing or to facilitate inter-firmware communication, timers of various kinds, and models that allow a peripheral to be associated with one or more interrupts (discussed in \S\ref{sec:interrupt-support}).
Finally, as some peripherals cannot be implemented using automated or generic approaches, but are essential to enable security analyses, e.g., complex I/O devices for communication over \ac{CAN}, we provide an observers-based API (\S\ref{sec:observers}) to \textbf{manually implement peripherals}. We demonstrate this in \S\ref{sec:rh850} by implementing \ac{CAN} network interface controller.

\subsubsection{Inferring Expected Peripheral Behavior}
\label{sec:peripheral-solving}

Our peripheral check solver can be applied in many rehosting scenarios, as firmware often only performs comparisons against a peripheral's status register to ensure it is initialized. These checks usually occur as tight loops whose exit condition depends on certain bits of a \ac{MMIO} register being set or clear, similar to \cref{lst:pchecking_c}. They generalize to the following: \pA the firmware reads the peripheral's \ac{MMIO} register, \pB it checks the value read using a boolean predicate, \pC if the predicate yields false, the loop repeats, otherwise execution continues past the check. We use this pattern to identify peripheral checking loops.

\noindent\begin{minipage}{\linewidth}
\vspace{1em}
\begin{lstlisting}[style=CStyle, label={lst:pchecking_c}, caption={Example of a peripheral checking loop.}, captionpos=b, escapechar=§]
do {
   mmioValue = read_4(0xfff81104);
} while ((mmioValue & 4) == 0);
\end{lstlisting}
\end{minipage}

\begin{figure}[t]
  \centering
  \resizebox{0.9\columnwidth}{!}{
    \def\svgwidth{0.8\textwidth}
    \large
    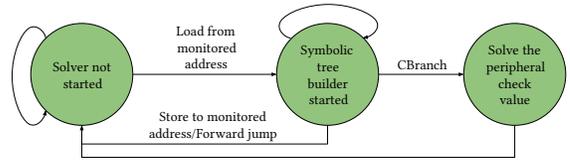

  }
  \vspace{-0.3cm}
  \caption{Peripheral model inference to bypass status checks. Our solver monitors loads from \ac{MMIO} register addresses and builds an expression tree of the constraints imposed upon the value read. Upon reaching a conditional (\texttt{cbranch}), it uses \ac{SMT} solver to compute an assignment for the register that allows us to exit the checking loop.
  \label{fig:prprsolver}}
  \vspace{-2em}
\end{figure}

\noindent\textbf{Bypassing status checks:} To compute \emph{satisfying} values to exit checking loops, we use localized symbolic execution, as shown in \cref{fig:prprsolver}. Our technique works by monitoring and symbolizing read accesses to our configured peripheral \ac{MMIO} region. For each access, we track its data-flow and build an expression tree capturing the constraints imposed upon the read value. Upon reaching a conditional branch that is dependent on our read value, we use a \ac{SMT} solver~\cite{DBLP:journals/jsat/NiemetzPB14} to find an assignment that enables us to explore the ``not taken'' branch target. We then execute the branch; if it leads us to re-enter of the checking loop, we replace the value read from the \ac{MMIO} register with the value computed using the solver.

\subsubsection{Universal Peripheral Models}
\label{sec:universal-peripheral-backend}

Building support for common peripherals for many different devices is both tedious and error-prone. Thus, it is desirable to build generic peripheral models instead. However, different \acp{MCU} use different \ac{MMIO} registers and bit masks to perform similar tasks, complicating the implementation process. For example, to control the start/stop of a \ac{CMT}, the C167CR \acp{MCU} use bit 6 at address \texttt{0xff42} \cite{c166ref}, while SH-2A \acp{MCU} use bit 0 at address \texttt{0xfffec000} \cite{sh2hardref}.
Our \emph{universal} peripheral models overcome this difficulty generically.

Within \toolname, we support \emph{universal} peripherals using two methods. \pA We provide an interface to intercept and redirect any function call, e.g., \ac{HAL} API calls. This enables \toolname to handle many kinds of I/O-based peripherals using generic handlers, without concern for the low-level hardware details. \pB We provide memory/register read/write hooking, which enables us to model MMIO-based peripherals generically. Our MMIO peripheral models implement data and status handling logic using a common interface that enables them to be used with our universal peripheral framework.
Our framework then allows us to configure each model declaratively by specifying the address, bit masks, and handler functions associated with a given peripheral's \ac{MMIO} registers. To facilitate model composition and architectural quirks, we provide a means to override each model's handler functions, which enables us to modify and combine their behaviors arbitrarily. \cref{lst:peripheral_example} demonstrates how the CMSTR register of SH2A \ac{CMT} can be configured using our generic \texttt{CompareMatchTimer} model.

\noindent\begin{minipage}{\linewidth}
\vspace{1em}
\begin{lstlisting}[style=rust-code, label={lst:peripheral_example}, caption={Configuration of a Compare \& Match Timer model using our universal peripheral backend.}, captionpos=b, escapechar=§ ]
let mut cmt = CompareMatchTimer::default();
cmt.map_function_addr_read(ADDR_CMSTR, 0x01,
                           &CMTFun::is_enabled);
cmt.map_function_addr_write(ADDR_CMSTR, 0x01,
                            &CMTFun::set_enable);
\end{lstlisting}
\vspace{-1em}
\end{minipage}

\subsection{Interrupt Support}
\label{sec:interrupt-support}

While many peripheral interactions occur synchronously, i.e., exclusively via \ac{MMIO}, others happen asynchronously. A common example is the \emph{tick} of a timer peripheral used in RTOS-based firmware to implement task scheduling. Such asynchronous events manifest in firmware via interrupts, which are handled by inducing a context switch to a so-called \ac{ISR}. This type of context switch is almost always facilitated by hardware, and works by first saving a snapshot of the current execution context to a vendor-defined area of memory, and then redirecting control-flow to the interrupt handler routine. To return from an \ac{ISR}, either the interrupt routine or the micro-controller itself will be responsible for restoring the execution context depending on the implementation of the vendor. These type of interrupts are commonly used in event driven firmware, where most functionality is implemented via \acp{ISR}. To handle this diversity in \toolname, we provide a generic interrupt handling framework, implemented using observers.

Similar to our universal peripheral backend (\S\ref{sec:universal-peripheral-backend}), our interrupt handling backend is also specification-based. It enables us to specify an interrupt by configuring handlers for its triggering behavior and restoration logic. Within \toolname, we provide a number of \emph{default} interrupt behaviors, e.g., to associate an interrupt trigger with a peripheral, and to perform context restoration based on specified registers or memory ranges. Our backend tracks each configured interrupt's \emph{enabled} and \emph{triggered} status, the \ac{ISR} or analysis-specific functionality to perform when it is triggered, and the context restoration logic to execute after it has been handled. We provide a complete example in \S\ref{sec:case_sh2a}.

\subsection{Integration with External Tools}\label{sec:inte-extern-tools}

As highlighted by Fasano et al.~\cite{conf/asiaccs/FasanoBMLBDEFLG21}, rehosting is an iterative process that requires human intervention and debugging. To that end, we integrate \toolname with widely used binary analysis and reverse-engineering tools, to reduce the manual effort involved.

\noindent\textbf{Trace visualization:} \toolname captures execution traces in a format supported by the widely used Binary Ninja and IDA Pro trace visualization plugin, Lighthouse~\cite{lighthouse}. As part of \toolname, we developed a Ghidra plugin providing comparable functionality. These plugins provide an intuitive view of the control-flow taken by our simulators, and can be used to quickly diagnose the cause of unexpected behavior and pinpoint the firmware regions responsible. An example is shown in \cref{sec:appdx:ghidra_trace}.

\noindent\textbf{Peripheral register name mapping: } When reverse engineering end-user firmware, we often encounter hundreds of functions that reference memory mapped peripheral registers. Unfortunately, tools such as Ghidra cannot automatically recover the names of all of these registers, and will often render them as raw addresses. When a firmware references many peripherals, this can be a significant hindrance. Thus, to aid our understanding of peripheral behavior when rehosting, we provide a peripheral register mapping plugin. Our plugin extracts a mapping of peripheral register names to addresses from C-header files distributed with vendor SDKs, which can be applied to Ghidra databases.

\noindent\textbf{Fuzzing and coverage reporting: } Fuzz testing is fast becoming a de facto means of discovering security flaws in many kinds of software. LibAFL~\cite{libafl} provides a robust base for constructing fuzz harnesses in a modular way. We use LibAFL's \texttt{Executor} trait to implement support for fuzzing firmware rehosted in our \acp{VXE}. We report coverage to LibAFL using an execution observer, and mimic \textsc{LAF-Intel}-style~\cite{lafintel} comparison splitting on our generated IR, which enables us to feedback finer grained coverage, without requiring \emph{a priori} recompilation or rewriting of binary-only targets.

\section{Evaluation}

We evaluate \toolname using three criteria: \pA performance, \pB implementation flexibility, and \pC real-world usability. For \pA we benchmark \toolname's raw execution performance, and compare it to two existing tools for firmware emulation: Unicorn and Ghidra's emulator. For \pA and \pC, we measure the extent our lifter is able to optimize its generated \ac{IR} for large \ac{ECU} firmware images, and compare the number of  emitted \ac{IL} operations to naive lifting without any optimization. We assess \pB and \pC together through six case-studies, which demonstrate using \toolname to perform different kinds of firmware analysis, including: inter-device analysis of two rehosted ARM firmware, inference of undocumented memory mapped peripheral interactions in a Renault \ac{BCM},  reverse engineering of the \ac{UDS} handler functions of a Land Rover Discovery's \ac{TCU}, fuzz testing to rediscover a backdoor in the ``Security Access'' \ac{UDS} service of a Volkswaggen Passat's \ac{IC}~\cite{conf/esorics/HerrewegenG18}, and emulation of the ThreadX RTOS implementation of a Range Rover Evoque \ac{TCU}. Across our case studies we demonstrate that \toolname is capable of enabling complex dynamic analyses built upon symbolic execution, fuzz testing, and micro execution~\cite{conf/icse/Godefroid14}.

We perform our experiments on a machine with a 32-core AMD Ryzen Threadripper 3970X and 128GB RAM.

\subsection{Performance Benchmarks\label{sec:perf_bench}}

We evaluate the performance of \toolname using two metrics: execution time and optimization of emitted IL operations. In the evaluation of execution time, we compare our tool with two existing emulation frameworks: Unicorn and Ghidra. We compare \toolname to Unicorn, as it is widely used for rehosting and performing security analyses~\cite{conf/wisec/MaierSP20,conf/woot/MaierRH19,conf/uss/RugeCGH20}, and to Ghidra's emulator, as it is used as the backend for \texttt{afl\_ghidra\_emu}, a fuzzer released by AirBus targeting esoteric architectures~\cite{airbus2021fuzz}. We use ARM programs for our head-to-head benchmarks based on the assumption that it is the most optimized embedded architecture all tools support.
%\footnote{Commit: 772558119af66269742fe4dcc45ec6000a5a6ea7}
%(we use statically linked Rust bindings with LTO enabled)

We perform four sets of experiments: \pA a micro-benchmark to evaluate the performance of each tool on various analysis tasks, \pB a deeper inspection of the differences in performance between Unicorn and \toolname performing different kinds of hooking, \pC an evaluation of the performance and trade-offs of performing IR optimization, and \pD the effects of IR optimization on the number of IL instructions emitted.

\subsubsection{Micro-benchmarks}\label{sec:micro-benchmarks}
\begin{figure}[t!]
\centering
\resizebox{\columnwidth}{!}{
\footnotesize
\def\svgwidth{0.675\textwidth}
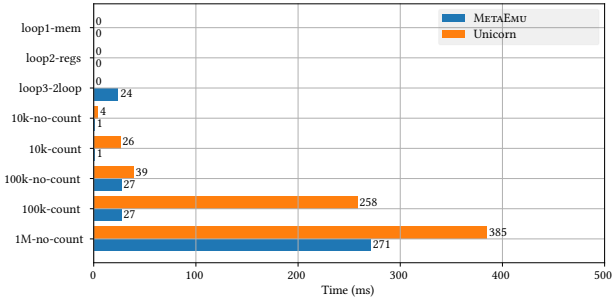
}
\vspace{-2em}
\caption{Performance comparison between \toolname and Unicorn.
%Times measured in milliseconds and averaged over 10k samples.
Tests 1-3 benchmark performance on tight loops updating state: registers or memory. Tests 4-8 benchmark programs of 10k, 100k, and 1M instructions. %Tests 9-13 benchmark the overheads of performing basic analyses using each framework: memory access hooking, function hooking, call context logging, and trace capture.
\label{fig:unicorn-benchmark}} %\todo[inline]{Maybe use the most representative benchmarks at here and put the rest in appendix? --agree. We could also try to float a table next to this graph for the IL optimization?}}
\vspace{-1.5em}
\end{figure}

We benchmark the performance of each tool on the programs described in Appendix~\ref{sec:benchmark-programs} and visualize the results in Figure~\ref{fig:unicorn-benchmark}. Our test harnesses measure execution time in milliseconds; we run each test 10,000 times and report the average time taken. We omit the results for Ghidra's emulator, as they are significantly worse than both tools shown---on average by at least an order of magnitude, e.g., ``1M-no-count'' takes 3234ms, compared to 271ms for \toolname and 385ms for Unicorn.

We find that \toolname performs favorably compared Unicorn on all but one benchmark.
We attribute the performance difference between \toolname and Unicorn on the ``loop3-2loop'' benchmark, to be due to the program being well suited to optimization by Unicorn's JIT compiler, as it contains a small number of basic blocks executed in two tight loops, where both loop bounds are hard-coded.

\subsubsection{Deeper Analysis of \toolname and Unicorn}

\begin{figure}[t!]
\centering
\resizebox{\columnwidth}{!}{
\footnotesize
\def\svgwidth{0.675\textwidth}
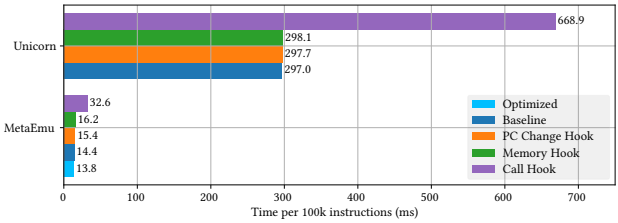
}
\vspace{-2em}
\caption{Comparison of \toolname and Unicorn whilst performing analyses using different kinds of hooks on a 100k instruction ARM-based sample program. We include two baselines for \toolname---one with IR optimization and one without---IR optimization gives a 4-5\% improvement.
\label{fig:hook-benchmark}}
\vspace{-1.5em}
\end{figure}

In this experiment, we benchmark the performance of \toolname and Unicorn while analyzing a sample program that consists of 100k instructions performing repeated memory transfer operations and function calls. In addition to baseline performance, we benchmark three different kinds of analyses: execution trace logging (\texttt{PC Change Hook}), memory access interception (\texttt{Memory Hook}), and function stubbing and call context logging (\texttt{Call Hook}). We include two baselines for \toolname---one with IR optimization and one without. We run each experiment 10,000 times and report the average time taken.
We do not include Ghidra's emulator in these experiments, as it does not provide sufficient hooking support and is dramatically slower than both \toolname and Unicorn.

%It worth mentioning that we use Unicorn for comparison in this evaluation because Ghidra Emulator does not have sufficient hooking support and its performance is clearly slower than both frameworks as shown in \cref{fig:optmi-bench}.

\smallskip
\noindent\textbf{Results: } We show the results of our benchmarks in \cref{fig:hook-benchmark}. Across all benchmarks, Unicorn is on average $\sim$20 times slower than \toolname. For \texttt{Memory Hook} and \texttt{PC Change Hook}, both \toolname and Unicorn have negligible performance overhead over baseline performance, while for \texttt{Call Hook}, both tools show larger overheads, which is due to the additional complexity of logging calling contexts. \toolname's IR optimizations provide an average of 4-5\% performance improvement.
Our benchmarks clearly demonstrate that our specification-based rehosting approach can perform at least as well as those based on Unicorn---without a performance trade-off---while also being much easier and flexible to extend.

\subsubsection{IR Optimization Performance and Trade-Offs}
In this experiment, we analyze the trade-offs of IR optimization by performing an extended analysis of the the ``loop3-2loop'' program from \S\ref{sec:micro-benchmarks} compiled for both ARM and C166 (omitted for Unicorn as it does not support C166). We measure the time each framework takes to execute the program's loop 1k times and 65k times, and measure the overheads of performing \toolname's IR optimizations both online and offline against a baseline without any optimizations enabled.

\smallskip
\noindent\textbf{Results: } \Cref{fig:optmi-bench} (Appendix~\ref{sec:ghidra-unicorn-metaemu}) shows the results of our benchmarks.
As mentioned in \S\ref{sec:micro-benchmarks}, Unicorn outperforms all frameworks on this benchmark, due to its JIT optimizer.
We find that although \toolname and Ghidra's emulator share the same processor specifications, \toolname is \textapprox 400\% faster on all benchmarks for both architectures.
\toolname performs most favorably when applying its optimizations offline, and when applying its optimizations \emph{online}, we only observe favorable performance on the longer running loop.  On the shorter loop, the overheads induced by optimization (C166: \textapprox15ms, ARM: \textapprox2ms) result in diminished performance. Thus, we find that the choice of whether to apply optimizations depends on the kind of analysis being performed and the size of the program being analyzed. We provide an option to disable optimizations as part of our synthesizer specification. As a rule of thumb, optimizations make most sense when running loop-based firmware, fuzzing, or when performing symbolic execution (due to the decreased size of symbolic formulas), and otherwise may negatively impact performance.

\subsubsection{Impact of Optimizations on Real Firmware\label{sec:optmi-reduction}}
\begin{figure}[t!]
\centering
\resizebox{\columnwidth}{!}{
%\footnotesize
\def\svgwidth{0.695\textwidth}
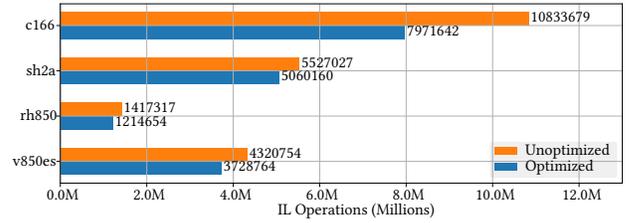
}
\vspace{-2em}
\caption{IR optimization performance for each \ac{ECU} firmware. Total number of operations calculated by lifting every basic block of each firmware; reduction calculated by number of eliminated operations. \label{fig:il-benchmark}}
\vspace{-1em}
\end{figure}

To evaluate the effectiveness of our optimizer on real-world firmware, we measure the total number of \ac{IL} operations before and after optimization when lifting every basic block in each of our \ac{ECU} firmware. The C166-based \ac{IC} firmware is \textapprox 512KB and contains 2609 functions, the RH850-based \ac{TCU} is \textapprox 32KB and contains 646 functions, the V850ES-based \ac{BCM} is 544KB and contains 2987 functions, the SH-2A-based \ac{TCU} is 8.4MB and contains 14,223 functions.

\smallskip
\noindent\textbf{Results: } We summarize \toolname's optimizer performance by \ac{IL} elimination in Figure~\ref{fig:il-benchmark}. Even though our results are a conservative approximation of our tool's real performance---accounting only for optimizations that lead to complete elimination of \ac{IL} operations---across all firmware we see reductions of hundreds of thousands of operations. For our C166-based firmware, for instance, we observe that \toolname reduces the number of operations by \textapprox 27\%: from \textapprox 10.8M to \textapprox 7.9M.

As our optimizations can be applied offline, they incur no overhead during simulation. However, when lifting, applying \ac{IR} optimization induces a non-negligible overhead of \textapprox 700$\mu$s per \ac{IR} block, compared to \textapprox 6$\mu$s without. This can be seen in \cref{fig:optmi-bench}, the optimization time is constant regardless of the number of iterations. We attribute the significant difference to our use of e-graphs and equality saturation~\cite{2021-egg} to find optimal \ac{IL} sequences---our optimizer effectively computes all simplifications of each \ac{IR} block and extracts the most optimal with respect to minimizing the IR size.

\subsection{Case Studies}\label{sec:case-studies}

In this section, we show how \toolname can be used to emulate and analyze firmware based on five different \acp{ISA}: ARM, C166, RH850, SH-2A, and V850E2---the latter four are not supported by any existing rehosting framework. We show that \toolname handles complex end-user firmware by rehosting four \ac{ECU} firmware, extracted from real automotive components. We also perform two other case studies: inter-device analysis (\S\ref{sec:inter-device}), and fuzzing and rehosting non-automotive firmware (\S\ref{sec:p2im}), using ARM-based firmware from the P$^2$IM data-set~\cite{conf/uss/FengML20}. The firmware from this data-set are simpler than our \ac{ECU} examples, and are chosen to enable us to more easily describe our experimental set-up and results.
We summarize our case studies in \cref{tbl:case-studies}.

\begin{table}
\vspace{-1em}
\caption{Summary of case studies performed. \toolname features evaluated: architecture support (\textbf{A}), inter/multi-device analysis (\textbf{M}), peripheral support (\textbf{P}), interrupt support (\textbf{I}), integration with external tools (\textbf{T}). Implementation size shows the approximate LoC for experiment (\textbf{E}) and specification for peripherals and execution mode (\textbf{S}). \label{tbl:case-studies}}
\centering
\footnotesize
\begin{tabular}{c | c | cc | ccccc}
\toprule
\multicolumn{1}{c|}{\multirow{2}{*}{\textbf{Firmware}}} & \multicolumn{1}{c|}{\multirow{2}{*}{\textbf{ISA}}} & \multicolumn{2}{c|}{\textbf{Impl. Size}} & \multicolumn{5}{c}{\textbf{Firmware Evaluated}} \\
\multicolumn{1}{c|}{} & \multicolumn{1}{c|}{} & \multicolumn{1}{c}{\textbf{E}} & \multicolumn{1}{c|}{\textbf{S}} & \multicolumn{1}{c}{\textbf{A}} & \multicolumn{1}{c}{\textbf{M}} & \multicolumn{1}{c}{\textbf{P}} & \multicolumn{1}{c}{\textbf{I}} & \multicolumn{1}{c}{\textbf{T}} \\

\midrule
Arduino \& NuttX firmware       & ARM       & 300  & 180 & \ding{51} & \ding{51} & \ding{51} & -         & \ding{51} \\
P2IM firmware (averaged)       & ARM       & 200  & 170 & \ding{51} & - & \ding{51} & -         & \ding{51} \\ % 14
Volkswagen \ac{IC}              & C166      &  260 & 100  & \ding{51} & -         & \ding{51} & -         & \ding{51} \\
Renault \ac{BCM}                & V850E2    & 160  & <10 & \ding{51} & -         & \ding{51} & -         & \ding{51} \\
Land Rover Discovery \ac{TCU}   & RH850     &  700$^\delta$ & - & \ding {51}& \ding{51}$^\ddagger$  & \ding{51} & - & \ding{51}\\
Range Rover Evoque \ac{TCU}     & SH-2A     & 370  & 110 & \ding{51} & -         & \ding{51} & \ding{51} & \ding{51} \\ %peripheral modeling(incl interrupt)
\bottomrule
\end{tabular}
\\
{
\vspace{3pt}
$\delta$: We manually implemented complex peripheral (CAN Bus) with hooking API

$\ddagger$: We rehost multiple instances of the firmware to perform fuzzing.
}
\vspace{-2em}
\end{table}

\subsubsection{Inter-Device Analysis}\label{sec:inter-device}

\begin{figure}[t!]
  \centering
  \resizebox{\columnwidth}{!}{
    \def\svgwidth{0.7\textwidth}
    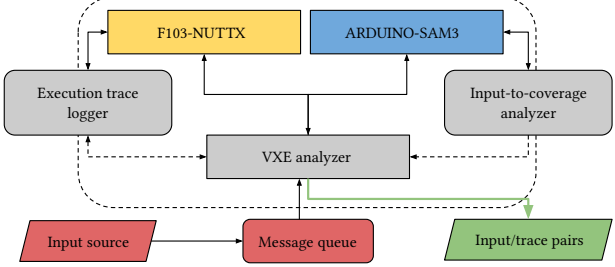
  }
  \caption{Inter-device analysis. Inputs are supplied outside the \ac{VXE} via a message queue, received by the F103, and forwarded to the ARDUINO. The F103 has an observer that logs the execution path corresponding to each input received and sent. The ARDUINO has an observer that analyzes the branch coverage corresponding to processing received inputs; on new coverage, it notifies the shared \ac{VXE} analyzer, which instructs the F103's observer to dump the execution trace matching the last sent input. The \ac{VXE} analyzer outputs input/trace pairs that yield new coverage.
  \label{fig:inter-device}}

 \vspace{-1.8em}
\end{figure}

In this case study, we demonstrate \toolname's capability to rehost multiple firmware in the same \ac{VXE} and perform an inter-device analysis. Each target firmware is ARM-based and adapted from the P$^2$IM~\cite{conf/uss/FengML20} data-set---one is based on the Arduino SDK, and the other on NuttX, as shown in \cref{fig:inter-device}.

\smallskip
\noindent\textbf{Objective: } The goal of this case study is two-fold: First, to show how two rehosted firmware can communicate, where one firmware (F103) is supplied input from outside the synthesized \ac{VXE}, and the other (ARDUINO) receives input from the other rehosted firmware (F103). Second, to demonstrate an analysis that captures execution traces from a rehosted firmware (F103) that leads to new branch coverage in another (ARDUINO). This case study models a typical scenario found in CAN networks, where one can only interact with a particular ECU by sending messages via another.

\smallskip
\noindent\textbf{Set-up: } We develop five observers. Three simulate peripherals: one to receive input via UART (outside to F103), one to transmit data via a TTY (from F103) and one to receive input via a serial port (into ARDUINO). Two perform analyses: an execution trace dumper that starts on a \emph{read} from UART and stops on a \emph{write} to a TTY, and a coverage tracker, which reports new branch coverage via the \ac{VXE}'s coordinator. The coordinator is configured to dump execution trace/input pairs that cause new branch coverage. We provide input to the F103 using random byte sequences, and specify a \emph{concrete execution mode} for each firmware.

\smallskip
\noindent\textbf{Discussion \& results: } Combined, our firmware configuration spans \textapprox 180 lines of code (not including processor specifications), while our observers take a further \textapprox 300. Our configuration accounts for overriding firmware functionality unrelated to our analysis task, creating a memory mapping for each firmware, and specifying the behavior of our analysis coordinator. Each rehosted firmware executes in parallel and faithfully simulates its expected behavior. This case study demonstrates how our approach can enable an otherwise tricky to reproduce analysis scenario with very little manual overhead. Further, it shows how \toolname can facilitate modeling peripheral and inter-device interactions, akin to those that regularly occur in real automotive networks, without requiring any hardware or source-code access.

\subsubsection{Volkswagen Passat Instrument Cluster (C166)}
\label{sec:c166}

In this case study, we demonstrate how \toolname can be used with LibAFL to rediscover a previously reported backdoor in the \ac{UDS} ``Security Access'' service of a Volkswagen Instrument Cluster~\cite{conf/esorics/HerrewegenG18}. The \ac{ECU} is based on the Infineon C166 architecture.
To support the C166 architecture, which is not currently distributed with Ghidra, we base our language definition on an open-source project~\cite{c166sla}; our definition consists of 1737 lines. We implement two extensions to our simulator: an intrinsic to handle the C166's \emph{DPP override} addressing~\cite{c166ref}, which enables the firmware to override which memory pages are accessed for a variable window of instructions, and an observer to implement its GPR bank switching.

\smallskip
\noindent\textbf{Objective \& set-up: } The backdoor is embedded in the \ac{UDS} handling routine. The routine receives two inputs: a buffer containing the \ac{UDS} request, which we populate with input from LibAFL, and a request ID, corresponding to a \ac{UDS} service, which we populate using the ID of the ``Security Access'' service (0x27). Authenticating against this service enables a client to access security critical services, e.g., ``Request Download'' (ID 0x34), which permits new software to be transferred to the \ac{ECU}. To authenticate, a client must complete a challenge-response handshake by sending a correct \emph{key} for a given \emph{seed}. The backdoor enables this check to be bypassed by supplying a specific hard-coded value in the request buffer (0xCAFFE012). The backdoor trigger is in the form of two 2-byte comparisons, and can easily be discovered by fuzzing. To do so, we configure our LibAFL harness to mark inputs that reach blocks corresponding to successful authentication attempts as if they induce a ``crash''.

\noindent\textbf{Discussion \& results: } We were able to trigger the backdoor with our fuzzer, but not as easily as expected. In fact, our initial attempt was unsuccessful, as the firmware relies on a timer peripheral to correctly execute the ``Security Access'' check---instead of validating our fuzzer's supplied ``key'', it enters an infinite polling loop.
On investigation, we found that this check first generates a challenge seed by repeatedly sampling from the timer's value register until a specific criterion is met.
Rather than bypassing the seed generation, we attached a peripheral to the firmware's simulator based on our generic \ac{CMT} model (\S\ref{sec:universal-peripheral-backend}). The specification takes just 6 lines; requiring masks for the enable bit, toggle on match bit, and current tick value, and the address range of the \ac{MMIO} of the timer peripheral. Our peripheral model emulates realistic behavior in our simulator when accessing the timer's memory mapped registers, \texttt{TxCON} and \texttt{Tx}, successfully enabling us to fuzz the firmware to discover the backdoor key.

\subsubsection{Renault Body Control Module (V850E2)}

In this case study, we use \toolname to aid in identifying peripheral access checks in firmware from a Renault \ac{BCM}. Even when an emulator is available for the architecture of a device, one of the major obstacles when performing rehosting is the lack of documentation for its peripherals. Usually, processor manuals provide the address range of \ac{MMIO} registers, but as each device can have a diverse array of peripherals, inferring the meaning of each address often requires manual reverse engineering. Unfortunately, when these documents are not in the public domain, we need to reverse engineer the \ac{MMIO} range itself to achieve even basic emulation. In this case study, we show how \toolname can help automate this task.

\smallskip
\noindent\textbf{Objective \& set-up: }
The objective of this case study is to facilitate basic emulation of our \ac{BCM} firmware, and identify peripheral access checks, without prior knowledge of specific peripheral register addresses. We use processor definitions adapted from those distributed with Ghidra. We obtain an approximate \ac{MMIO} peripheral address range from the processor manual of a related \ac{MCU} (V850E2/Fx4-G), as the peripheral manual for the \ac{MCU} (V850E2/Dx4) used in the \ac{BCM} is not publicly available.

\smallskip
\noindent\textbf{Identifying peripheral status checks:}
We perform our analysis in two steps. We first use our peripheral check solver to identify peripheral checking loops using registers mapped to the \ac{MMIO} range of a similar \ac{MCU} (\texttt{0xff400000}--\texttt{0xffff8000}). This enables us to identify and bypass some peripheral checks, however, our firmware still \emph{stalls} prior to reaching its main loop. To overcome this, we gradually widen our assumed \ac{MMIO} range by specifying larger bounds in our solver's configuration. After a few iterations, we determined that the real \ac{MMIO} range of the \ac{MCU} is much wider than expected: we discovered checks in the range \texttt{0xffff8000}--\texttt{0xffffffff}. By reconfiguring our peripheral check solver to use this extended range, we were able to bypass the initialization checks performed by the firmware. This case study thus demonstrate that \toolname's specification-based approach can enable analyses of this kind without developing custom peripheral models, or  device-specific heuristics.

\subsubsection{Land Rover Discovery Telematics Unit (RH850)}
\label{sec:rh850}

In this case study, we use \toolname to rehost the \ac{TCU} from a 2018 Land Rover Discovery in order to reverse-engineer its \ac{UDS} handler routines. To aid our reverse-engineering, we use fuzzing to identify \emph{valid} \ac{UDS} requests. The \ac{ECU}'s firmware is based on the RH850 architecture. It's worth mentioning that despite the RH850 and V850 are from similar microcontroller family, there are still differences in instruction set and the peripheral layout.

\smallskip
\noindent\textbf{Objective \& set-up: } \acp{TCU} are responsible for communicating with and monitoring other \acp{ECU} on the \ac{CAN} bus, and reporting diagnostics data and metrics over other mediums, such as LTE, hence have a large attack surface. \ac{UDS} is the protocol responsible for transmitting diagnostic information and is implemented on top of \ac{CAN}. Among other functionality, \ac{UDS} can be used to re-flash or dump an \ac{ECU}'s firmware. The goal of our analysis is to reverse engineer the UDS handling functions in our target firmware. These functions take input from a CAN peripheral, process it, and act accordingly. To analyze them, we take a four-step approach:

\begin{enumerate}
    \item Since \ac{UDS} is a \ac{CAN}-based protocol, to determine how to use it to communicate with the device, we first need to identify the device's \ac{CAN} IDs.
    \item Next, to interface with the device's rehosted firmware over \ac{UDS}, we create a model for its CAN peripheral.
    \item As the firmware interfaces with many peripherals unrelated to our analysis task, yet relies on them being successfully initialized to configure its internal state prior to executing its main loop,
    we attach a peripheral check solving observer to supply suitable values when accessing their \ac{MMIO} registers.
    \item Finally, to identify valid \ac{UDS} requests, we use a LibAFL-based fuzzer to supply input to the firmware via its \ac{CAN} peripheral, and track which inputs lead to new coverage.
\end{enumerate}

\noindent\textbf{CAN ID identification: }
In a modern automobile, the \ac{CAN} bus is the standard medium for \ac{ECU}s to communicate with each other and the outside world. Thus, it is usually the first interface/input source we investigate when analyzing automotive firmware. On the \ac{CAN} bus, each \ac{ECU} is assigned one or more CAN IDs, and will only respond to messages sent to those IDs. As many manufacturers attempt to keep CAN IDs secret, it is a non-trivial task to identify them effectively without analyzing the \ac{ECU}'s firmware.

As mentioned previously, automotive microcontrollers usually interface with their peripherals and, thus, the \ac{CAN} bus, by reading and writing to \ac{MMIO} registers. Each such register controls either a peripheral's behavior or act as an input/output buffer. From our experience, for peripherals that interface with the \ac{CAN} bus, one register will contain the \emph{listening} \ac{CAN} ID. Hence, if we know the address this register is mapped to, we can use it to recover the IDs the \ac{ECU} \emph{listens} on by monitoring values written to it.

Unfortunately, exhaustively enumerating all execution paths that write to the register requires near complete emulation of the entire firmware---defeating the purpose of rehosting. Flood execution~\cite{conf/raid/WilhelmC07}, however, provides a means to approximate all execution paths in a bounded manner, and the micro execution-based variant proposed by Gotovchits et al.~\cite{conf/bar/gotovchits2018} provides a means to perform flood execution in the presence of failing memory accesses. To recover \ac{CAN} IDs, we use \emph{flood execution} mode with our synthesized simulator to enumerate paths, and use an observer to log writes to the register documented to store \ac{CAN} IDs. Through this process we recover three \ac{CAN} IDs: \texttt{0x7df}, \texttt{0x18db33f1}, and \texttt{0x700}.

\begin{figure}[t]
  \centering
  \resizebox{\columnwidth}{!}{
    \def\svgwidth{0.9\textwidth}
    \Large
    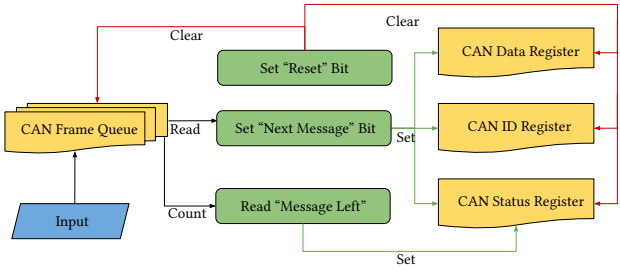

  }
  \vspace{-1.5em}
  \caption{Data-flow diagram of RH850 \ac{CAN} peripheral. We store incoming \ac{CAN} messages in a FIFO queue (blue box). When the firmware sets the ``next message'' bit in its CAN MMIO register, we dequeue a CAN frame, use it to populate the \ac{CAN} data and \ac{CAN} ID registers, and then set the data available bit in the CAN status register. Upon a \textit{reading message left} event, we update the ``message left'' counter in the CAN status registers with the length of the queue. When the firmware sets the peripheral's ``reset'' bit, we clear the queue and reinitialize the registers.   \label{fig:canperi}}
  \vspace{-2em}
\end{figure}

\smallskip
\noindent\textbf{\ac{CAN} peripheral modeling: }
We model the device's \ac{CAN} peripheral using observers, as shown in \cref{fig:canperi}, and interact with our model via SocketCAN~\cite{socketcan}. We use an observer to hook \emph{reads} and \emph{writes} to the \ac{CAN} registers to simulate receiving and sending \ac{CAN} frames. When sending data, we build \ac{CAN} frames using values from the CAN status register (\texttt{TMSTS}), the receive ID register (\texttt{RFID}) and receive data register (\texttt{RFDF}); after transmission, we set the firmware's transmission success bit. When receiving data, we read a frame from our input source, and use it to set the following registers: \texttt{TMC}, \texttt{TMID}, \texttt{TMPTR}, and \texttt{TMDF}. %Then we are able to communicate with the firmware using this peripheral.

\smallskip
\noindent\textbf{Bypassing peripheral status checks: } To bypass the initialization checks of the device's other peripherals, we use our symbolic status check solver (\S\ref{sec:peripheral-solving}). We configure it using the full peripheral address range: \texttt{0xFF400000}--\texttt{0xFFFFAFFF}. Alongside our CAN peripheral, this enables us to execute the firmware from reset to its \ac{UDS} request processing loop.

\smallskip
\noindent\textbf{Fuzzing for valid \ac{UDS} requests: } To fuzz for valid \ac{UDS} requests, we use \emph{concrete execution} mode to run our rehosted firmware from reset to the \ac{UDS} function, and then use an observer to take a snapshot of the simulator's state.
We build a LibAFL-based harness which starts execution from our snapshot and supplies input to the \ac{UDS} routine via our \ac{CAN} peripheral. We implement a custom \texttt{EventManager} for LibAFL that supports sharing coverage and \emph{interesting} inputs across multiple \toolname simulators via our \emph{analysis coordinator}. This enables us to rehost many instances of our firmware and fuzz the \ac{UDS} routine in parallel. Through our analysis, we generated inputs that explored execution paths covering 898 unique \ac{IL} branches, which uncovered 8 \ac{UDS} request handlers.

\subsubsection{Range Rover Evoque Telematics Unit (SH-2A)}\label{sec:case_sh2a}

In this case study, we demonstrate how \toolname can be used to rehost a Range Rover Evoque \ac{TCU} SuperH-2A firmware. The firmware is based on ThreadX RTOS, which executes multiple tasks concurrently. To rehost, it requires robust peripheral models to correctly perform task switching---something that cannot be achieved using simpler symbolic modeling approaches, such as the peripheral check solver used in our other case studies.

\noindent\textbf{Objective \& set-up: } Through manual reverse engineering, we found that the firmware uses a timer interrupt to facilitate task scheduling, which is performed by a \ac{CMT} peripheral. The goal of this case study is: First, to show how \toolname facilitates implementing a complex peripheral that can faithfully trigger task switches. Second, to show how \toolname can emulate hardware-supported multi-tasking.

\begin{figure}[t]
  \centering
  \resizebox{0.86\columnwidth}{!}{
    \def\svgwidth{0.68\textwidth}
    \large
    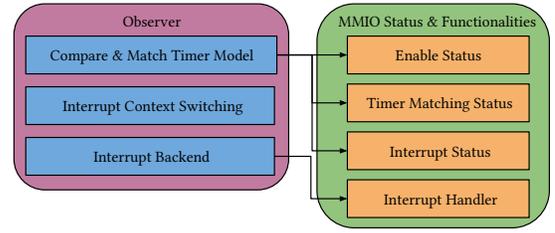
  }
  \vspace{-1em}
  \caption{Overview of our SH-2A \ac{CMT} model, implemented using an observer and our universal peripheral and interrupt backends. Our \ac{CMT} model is responsible for tracking the enabled, timer matching, and interrupt status of the peripheral. We use our interrupt backend to specify a custom interrupt handler and ensure the PC and SP registers are preserved during the context switch.
  \vspace{-1.5em}
  \label{fig:timer-interrupt}}
\end{figure}

\noindent\textbf{Discussion \& results: } Since the device's \ac{CMT} peripheral acts as both a timer and interrupt source, we can implement it using a combination of our generic \ac{CMT} model (\S\ref{sec:universal-peripheral-backend}) and our interrupt backend (\S\ref{sec:interrupt-support}), as shown in \cref{fig:timer-interrupt}.
To configure our \ac{CMT} model, we specify the mapping between the peripheral's registers and the model's set/unset actions. This enables our model to update and track its enabled status, interrupt status, and timer matching status. To simulate the timer's counting and interrupt triggering behavior, we implement a simple observer that increments the timer's value by one each time a new architectural instruction is lifted, and \emph{fires} an interrupt using \toolname's interrupt backend. Our implementation (optimistically) assumes that
each instruction takes one clock cycle, and the timer increases based on the device's clock.
We configure our interrupt backend to trigger a jump to the firmware's timer \ac{ISR} when it is instructed to \emph{fire} by our timer observer. Since the context switch is normally performed in hardware, we configure \toolname to preserve the status register \texttt{SR} and program counter \texttt{PC}
prior to jumping. The firmware uses the
\texttt{rte} instruction to return from interrupt handlers, and its logic is implemented in the architecture's processor specification.
Our peripheral requires just \textapprox 250 lines of specification and code. To test our peripheral model, we attach an execution tracer observer to our simulator and run the device from reset. Our traces show that the emulated firmware correctly performs task switches and mirrors the real firmware's behavior.

\subsubsection{P$^2$IM dataset firmware}\label{sec:p2im}

The P$^2$IM authors~\cite{conf/uss/FengML20} provide a data-set consisting of 10 ARM-based firmware based on open-source projects, which all use a \ac{HAL} to interact with their peripherals. In this case study, we use the \texttt{Gateway} and \texttt{Soldering\_Iron} firmware to demonstrate the correctness of \toolname and show that it can achieve the same analysis outcomes as existing rehosting frameworks. We perform two experiments: \pA we fuzz test the \texttt{Gateway} firmware, and \pB we use the \texttt{Soldering\_Iron} firmware to test \toolname's support for handling tricky (DMA-based) peripherals.

\textbf{Set-up:}
For experiment \pA, we set up a basic fuzz harness to supply I2C query packets by hooking the HAL of the firmware. Similar to past work~\cite{muench2018you}, we mark part of the firmware's memory as a ``red zone'' (write permission disabled) to perform sanitization and detect any memory corruption bugs. For experiment \pB, we attempt to test if \toolname can run the firmware from its entry-point, allow it to perform peripheral initialization, and execute its main loop without crashing due to peripheral mishandling.

\textbf{Discussion \& results:}
For \pA, we were able to trigger an out of bounds write with inputs exceeding a length of 62 bytes (the first two bytes of \texttt{i2cRxData} are populated using \texttt{i2c\_device\_info.addr} and \texttt{i2c\_device\_info.reg}) after only three iterations of our fuzz harness. This demonstrates that \toolname is usable for analyzing firmware beyond our original use-case of automotive firmware, as well as its suitability for fuzz testing.

For \pB, the author's of P$^2$IM~\cite{conf/uss/FengML20} describe two \emph{false} crashes/hangs induced by their framework when rehosting the \texttt{Soldering Iron} firmware. The first is caused by misclassification of a peripheral register, and the second is due to P$^2$IM missing support for DMA-based peripherals. As \toolname can support peripherals generically, we were able to successfully rehost this firmware by intercepting DMA interactions by the firmware's HAL interface in a manner similar to past work by Clements et al.~\cite{conf/uss/ClementsGSGFKVB20}. We thus avoid the false crashes experienced by P$^2$IM and $\mu$Emu~\cite{conf/uss/Zhou21}. Since \toolname provides an API to intercept any memory read/write (i.e., \texttt{hook\_memory\_write/read}), it can be easily extended to provide DMA support similar to DICE~\cite{conf/sp/Mera21}. We note that since the author's of P$^2$IM do not provide source code for this firmware\footnote{\url{https://github.com/RiS3-Lab/p2im-real_firmware/issues/2}}, we had to manually stub out the functionality relating to AFL, e.g., the call to \texttt{aflCall}, as it was unnecessary for our experiment. %We opted not to recompile the firmware ourselves, as many of the P$^2$IM firmware appear to be heavily modified\footnote{\url{https://github.com/RiS3-Lab/p2im-real_firmware/issues/3}}.

% As claimed by P2Im only 1/10 in their test set used DMA
% \todo{Discuss about coverage}
%\todo{Discuss the False crash in soldering Iron Firmware:
%
%P2IM paper, they described the false crash due to CR being false catagrized as DR. However in our tool, we are not using register catagrize approach to handle peripherals so it will not happen in our peripheral handding approcah. Besides, in terms of fuzzing, we are using execution hooks to hook HAL calls and feed in fuzzing data in our test. So it will also not happen with out tool.
%DMA false crash, we provide toolset that allow DICE integration. }

\section{Human effort}
In this section, we discuss the human effort involved in using \toolname for rehosting.
As with any kind of reverse engineering, rehosting necessarily requires manual intervention for tasks that cannot be automated---from implementing complex peripherals for I/O to identifying functions to hook and override. \toolname attempts to reduce this manual effort by moving towards a specification-based approach from one that is purely programmatic.
In \cref{tbl:case-studies}, we show the human effort involved in each of our case studies in terms of lines of code---all require under 1kloc---even in the case of complex inter-device analysis. Concretely, each case study took the author's less than one day to implement.

The advantage of using our approach is twofold: firstly, a spec\-ifi\-cat\-ion-based approach provides a DSL which enables much faster iteration of manual tasks such as implementing missing \ac{ISA} instructions or peripheral models. As we separate specification files from the implementation of our tool, we can prototype new features without recompiling the tool, something not possible with QEMU-based rehosting approaches. Secondly, by using the same \ac{ISA} specification language as Ghidra, we benefit from the maintenance and testing efforts of a large and active community. Ghidra's repository contains specifications for 28 architectures and when combined with community projects~\cite{c166sla}, \toolname can rehost most \acp{ECU} without the need to implement any architectural support.

% While such an approach does not overcome the need for expert domain knowledge, e.g., to digest architecture documentation and implement \ac{ISA} or peripheral models, it does enable much faster iteration of those manual tasks. In particular, with our approach we can implement new or missing \ac{ISA} instructions by modifying a configuration file using a domain specific language suited for the task, which can then be used by \toolname without any recompilation. This is a stark contrast to existing rehosting frameworks, such as those that rely on QEMU or Unicorn, which need to be recompiled to implement or modify instruction behavior---incurring a significant  time overhead and hindering the rehosting process. Meanwhile, with \toolname, we avoid such overheads and maintain comparable runtime performance.

\section{Limitations}

In this section, we provide a discussion of \toolname's limitations.
The correctness of \toolname's lifter depends on the processor definitions it uses as input. Fortunately, as they are based on Ghidra's language definitions, we benefit from Ghidra's large and active community that regularly contributes fixes.

The three peripheral interfaces in \toolname provide a simple way to implement peripherals for automotive microcontrollers of atypical architectures, however, we have less peripheral models for common microcontrollers such as ARM or MIPS based devices. This makes it difficult to produce a direct comparison with existing emulators such as QEMU. As future work, we intend to explore adding a compatibility layer to allow \toolname to use QEMU-derived peripheral models to address this issue, however, it would require implementing QEMU's Object Model, memory, sysbus, interrupt, and peripheral APIs in \toolname to do so.
Our universal peripheral backend (\S\ref{sec:universal-peripheral-backend}) does not fully support \ac{DMA}-based I/O, except when firmware interfaces with it using a \ac{HAL}.
Support for peripherals requiring this, however, can easily be added using \toolname's execution observers (\S\ref{sec:observers}). %We leave complete support as future work.

While \toolname's runtime performance is comparable to Unicorn, there are possibilities for improvement, such as adding a JIT compiler for our \ac{IL}. We believe such an enhancement will lead to even greater performance for our specification-based approach. However, it is unclear how to generically implement a JIT compiler that supports all of \toolname's different execution modes (\S\ref{sec:execution-modes}).

\section{Discussion \& Related Work}
Approaches to firmware rehosting cover a broad spectrum: in terms of the kind of analyses they facilitate, the degree to which they rely on hardware, the kinds of device they support, and the fidelity and faithfulness of the environments they transplant firmware into.

\textsc{Avatar}~\cite{conf/ndss/ZaddachBFB14,conf/bar/muench2018} and \textsc{Surrogates}~\cite{conf/woot/KoscherKM15} propose hardware-in-the-loop analysis, which enables a device to be analyzed without handling many of the complexities of its peripherals. It permits a kind of hybrid methodology where a fast host can emulate most of the firmware and rely on the real device for peripheral I/O and interrupt handling. Many techniques have capitalized on these seminal works; for instance, Ruge et al.~\cite{conf/uss/RugeCGH20} use a hardware-in-the-loop approach to fuzz for vulnerabilities in bluetooth chips, while Gustafson et al.~\cite{conf/raid/GustafsonMSRMFB19} use such an approach as a basis for their tool, Pretender, which automatically infers peripheral models from execution traces and device I/O behavior. \textsc{Inception}~\cite{conf/uss/CorteggianiCF18} and \textsc{HardSnap}~\cite{conf/dsn/CorteggianiF20} use a hardware-in-the-loop approach for their debugger component; both attempt to handle the nuances of complex firmware and devices with multiple peripherals under symbolic execution.

In contrast to hardware-dependent approaches, emulation-based approaches do away with hardware interaction altogether, and to varying degrees attempt to emulate a device and its peripherals. Of the approaches that make their implementations open-source, we observe that almost all rely on either QEMU (e,g.,\cite{conf/uss/HarrisonVPSG20}) or Unicorn (e.g.,\cite{conf/wisec/MaierSP20}) as the basis for their \ac{VXE}.
Clements et al.~\cite{conf/uss/ClementsGSGFKVB20} emulate peripherals by hooking \ac{HAL} APIs provided by many vendor SDKs. They simulate peripheral interactions through generalized models that receive input and supply output via the hooked \ac{HAL} functions. In~\cite{conf/bar/clements21} they extend their approach to support VxWorks-based devices. Feng et al.~\cite{conf/uss/FengML20} rehost firmware to facilitate fuzz testing. To handle peripherals, they learn appropriate values for \ac{MMIO} peripheral interactions based on device-specific \emph{abstract peripheral models}. Mera et al.~\cite{conf/sp/Mera21} propose a method to handle DMA-based peripheral inputs, similarly targeted at fuzzing rehosted firmware. Liu et al. ~\cite{firmguide} use model-guided execution to generate QEMU peripheral models from kernel device-tree and source code. Cao et al.~\cite{conf/acsac/CaoGM020}, Johnson et al.~\cite{conf/uss/JohnsonBZMCSL21}, and Zhou~\cite{conf/uss/Zhou21}, all leverage symbolic execution to learn \emph{satisfying} values to bypass peripheral checks.
Hernandez et al.~\cite{bhshannon20} achieve full-system emulation of closed-source Shannon baseband firmware by adding missing architectural and peripheral support in QEMU, they later demonstrate that such an approach can be extended to other basebands~\cite{conf/ndss/hernandez22}.
In contrast to the aforementioned approaches, Milburn et al.~\cite{alyssa2018glitch} build a custom emulator and peripheral models to rehost an automotive instrument cluster; they use their emulator to aid in reverse-engineering the firmware's \ac{UDS} commands.
Meanwhile, Davidson et al.~\cite{conf/uss/DavidsonMRJ13} use full-system symbolic execution to discover vulnerabilities in PIC32 devices; they propose using specifications to configure interrupt and peripheral memory mappings, allowing their approach to be adapted to analyze different device configurations.

 While pure emulation and hardware-in-the-loop approaches can achieve near complete support for a given device or family of devices, when based on commodity emulators, they are difficult to adapt to support devices based on esoteric architectures, such as automotive components, due to the substantial engineering effort required. Mera et al.~\cite{conf/sp/Mera21} highlight this difficulty in their evaluation---to test their approach on both ARM and MIPS32-based devices, they need to build separate prototypes of their tool for two different forks of QEMU, as neither variant supports both architectures. Hernandez et al.~\cite{conf/ndss/hernandez22} note the current impossibility of porting their baseband rehosting framework to work with Qualcomm basebands, due to lack of architecture support in the PANDA~\cite{conf/acsac/Dolan-GavittHHL15} QEMU fork.

As discussed in our evaluation, Ghidra provides a P-Code-based emulator, which can facilitate basic analysis tasks, including basic instruction hooking and manipulation of memory and register values. However, it lacks more advanced features for firmware reverse engineering, such as symbolic execution, and peripheral support. Although it is possible to add such functionality using its hooking API, its emulation performance is insufficient to support intensive tasks, such as fuzzing. Further, adding functionality to support dynamic changes to addressing modes, e.g., to correctly emulate C166-based firmware, would require extensive changes to the core of Ghidra's emulator framework---a significant engineering task.

In Table~\ref{tbl:comparison} (Appendix~\ref{sec:comparison-table}), we provide a feature, architecture, and analysis-support comparison of \toolname with the state-of-the-art, using the framework classification proposed by Fasano~\cite{conf/asiaccs/FasanoBMLBDEFLG21}.
The key difference between \toolname and the frameworks listed, is that our approach generically enables analysis of firmware not currently supported by other frameworks with little effort. Moreover, like \textsc{Avatar}, it is a general framework to build analysis tools, as opposed to a method to enable a specific type of analysis, e.g., fuzzing. As demonstrated through our case studies (\S\ref{sec:case-studies}), we can use \toolname to build analyses similar to those proposed by other approaches (e.g., \cite{conf/acsac/CaoGM020,conf/uss/JohnsonBZMCSL21}) in a completely architecture-independent way, without resorting to implementing those techniques for many different emulators, or limiting the approach to a few architectures.
As with other rehosting approaches, such as P$^2$IM, \toolname's peripheral support will lead to better coverage when performing analysis tasks, such as fuzzing, over an emulator with no peripheral support.

\section{Conclusion}
To conclude, we have presented the first architecture-agnostic framework capable of rehosting multiple devices simultaneously. Our \ac{IR} lifter and simulators provide fully generic, extensible, architecture support, and our universal peripheral models and peripheral solver enable simulation of realistic execution environments. We also have tight integration with binary analysis tools to help manual analysis. Through our case studies, we have demonstrated that our tool is flexible, efficient, and can drive the analysis of real-world automotive firmware whose architectures are not supported by existing state-of-the-art rehosting approaches.

%\balance
\section*{Acknowledgments}
This research is partially funded by the Engineering and Physical Sciences Research Council (EPSRC) under grant EP/R012598/1, EP/R008000/1 and EP/V000454/1.

\bibliographystyle{ACM-Reference-Format}
\bibliography{bibliography}

%\newpage
\appendix

\section{Execution Trace visualised in Ghidra}\label{sec:appdx:ghidra_trace}
\begin{figure}[H]
    \centering
    \includegraphics[scale = 0.2]{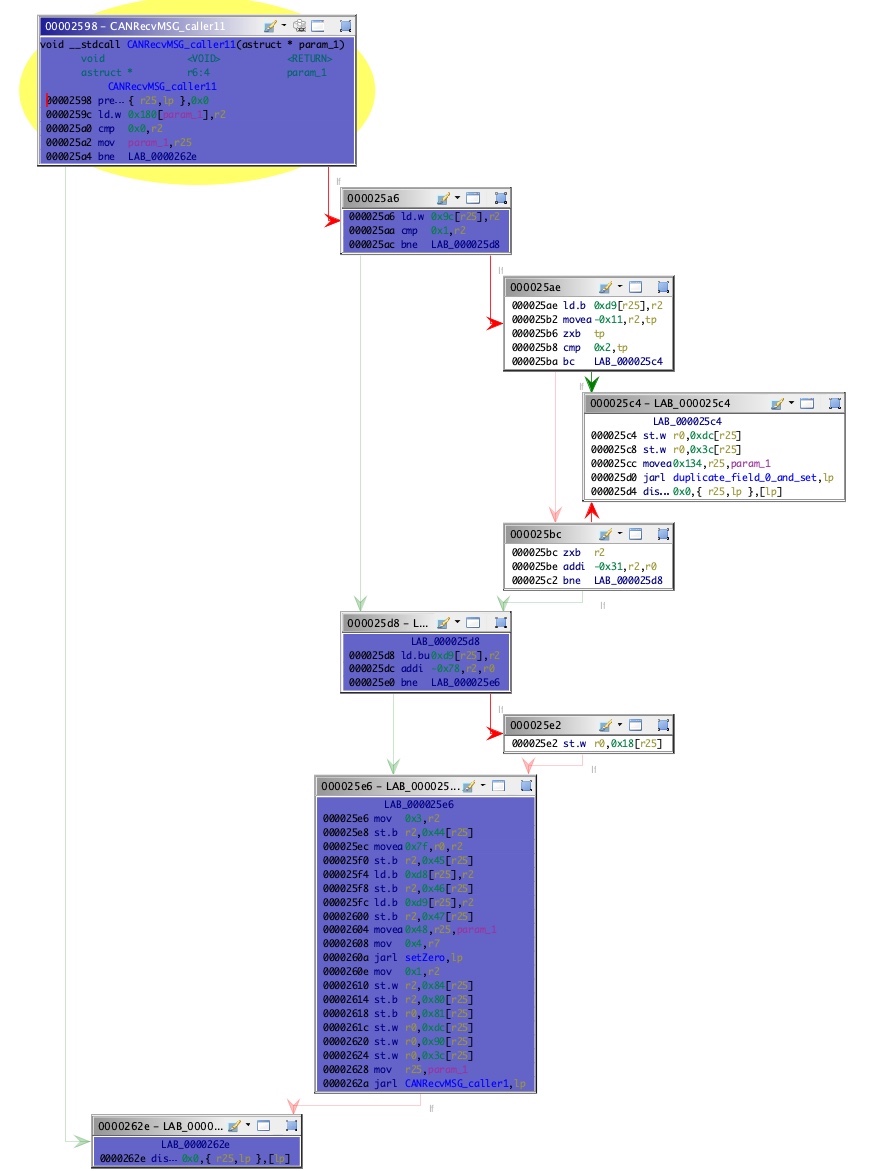}
    \caption{Example Function Graph processed with trace visualizer. The executed flow is colored with purple}
    \label{fig:ghidra_trace_color}
\end{figure}
\section{Execution Observer Event API}\label{sec:observer_api}
%\begin{table*}[h!]
\begin{center}
\footnotesize
\begin{tabular}{p{3cm} | p{4.5cm}}
\toprule
Name & Purpose \\
\midrule
hook\_operand\_read & Intercept a register or memory read. \\
hook\_operand\_write & Intercept a register or memory write. \\
hook\_memory\_read & Intercept a memory read. \\
hook\_memory\_write & Intercept a memory write. \\
hook\_register\_read & Intercept a register read. \\
hook\_register\_write & Intercept a register write. \\
hook\_architectural\_step & Intercept a program counter change and the lifted instructions at the new address. \\
hook\_operation\_step & Intercept a single IL operation step. \\
hook\_cbranch & Intercept a conditional branch. \\
hook\_call & Intercept a function call. \\
\bottomrule
\end{tabular}
\end{center}
%\end{table*}

\section{Benchmark Programs}\label{sec:benchmark-programs}
\begin{center}
\footnotesize
\begin{tabular}{p{2.3cm}@{\hspace{3pt}}| p{5.2cm}}
\toprule
Name & Description \\
\midrule
loop1-mem & Fibonacci sequence computed in a tight loop with a large N; multiple small blocks; uses stack to store all computed values. \\
loop2-reg & Fibonacci sequence computed in a tight loop with a large N; multiple small blocks; uses registers to store all computed values. \\
loop3-2loop & loop1-mem wrapped in a loop of 1k iterations: two tight loops. Maximizes optimization potential for Unicorn's JIT. \\
10k-no-count & 10k instructions (multi-block), no execution bound set.\\
10k-count & 10k-no-count with a counter updated after each processed instruction. We implement counting for Unicorn by hard-coding parameter 4 of \texttt{start\_emu} to be the number of instructions to process (i.e., 10k). \\
100k-no-count & 100k instructions (multi-block), no instruction bound. \\
100k-count & 100k instructions (multi-block), with a counter update after each instruction, 100k instruction bound.\\
1M-no-count & 1M instructions (multi-block), no instruction bound. \\
\bottomrule
\end{tabular}%\onecolumn
\end{center}

\section{ARM and C166 loop-based benchmarks}\label{sec:ghidra-unicorn-metaemu}

\begin{figure}[h!]
\centering
\resizebox{\columnwidth}{!}{
\footnotesize
\def\svgwidth{0.675\textwidth}
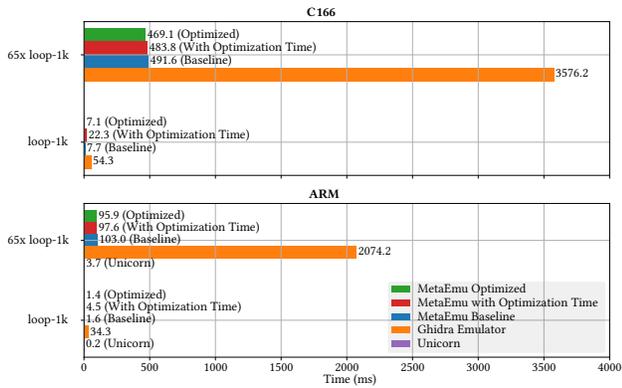
}
\vspace{-2em}
\caption{We benchmark Ghidra and \toolname's performance for both ARM and C166, and Unicorn for ARM, as it does not support C166. \toolname's optimization overhead is constant for both architectures on both test configurations---$\sim$15ms for C166 and $\sim$2ms for ARM. As a result, optimization is more useful when running the program for a large number of iterations (e.g., fuzzing).\label{fig:optmi-bench}}
\vspace{-1.5em}
\end{figure}

\onecolumn
\section{Comparison of \toolname with the state-of-the-art}\label{sec:comparison-table}

\begin{table*}[h!]
  \caption{A comparison of \toolname with the state-of-the-art.
  Layer refers to how the framework handles a rehosted device and its firmware: $\varepsilon$--untouched, $\alpha$--passed through, $\beta$--emulated, $\phi$--replaced, $\chi$--symbolic model.
  Type refers to the kind of devices the framework can rehost; we use the classification proposed by Fasano et al.~\cite{conf/asiaccs/FasanoBMLBDEFLG21}: 3--bare-metal (no OS), 2--embedded-device specific OS (e.g., Zephyr), 1--general purpose OS retrofitted for embedded firmware (e.g., Linux).
  Peripheral classifies how peripherals are handled: \ding{108}--hardware, \ding{119}--approximated (symbolic or inferred) \ding{109}--modeled. Multi-device denotes if the framework can rehost multiple targets simultaneously. Analysis support refers to the kinds the framework analyses enables.
  \label{tbl:comparison}}
  \vspace{-13pt}
  \centering
  \footnotesize
\begin{tabular}{@{\hspace{2pt}}c@{\hspace{3pt}}| c|c|c|c | c|@{\hspace{3pt}}c@{\hspace{3pt}}| c |@{\hspace{2pt}}c@{\hspace{2pt}}|@{\hspace{3pt}}c@{\hspace{2pt}}}
  \toprule
  \multirow{2}{*}{\bf Framework} & \multicolumn{4}{| c |}{\bf Layer}                                               & \multicolumn{2}{| c |}{\bf Target} & \multirow{2}{*}{\bf Peripheral} & \multirow{2}{*}{\bf Multi-device} & \multirow{2}{*}{\bf Analysis Support} \\
                                 & \bf Hardware          & \bf OS                & \bf Application       & \bf Function            & \bf Type  & \bf ISA(s)                     &                             &                                   \\
  \midrule
  \textsc{Avatar}~\cite{conf/ndss/ZaddachBFB14}  & $\alpha$/$\beta$/$\chi$ & $\varepsilon$/$\alpha$ & $\varepsilon$/$\alpha$ & $\varepsilon$/$\alpha$/$\chi$ & 2, 3 & ARM & \ding{108} & \hspace{4pt}\ding{51}$^\ddagger$ & Generic \\
  \textsc{Avatar}$^2$~\cite{conf/bar/muench2018}  & $\alpha$/$\beta$/$\phi$ & - & $\varepsilon$/$\alpha$ & $\varepsilon$/$\alpha$  & 3 & ARM & \ding{108} & \hspace{4pt}\ding{51}$^\ddagger$ & Generic \\
  DICE~\cite{conf/sp/Mera21}  & $\beta$/$\phi$ & $\varepsilon$ & $\varepsilon$ & $\varepsilon$ & 2, 3 & ARM, MIPS & \ding{119} & \ding{55} & Fuzzing \\
  HALucinator~\cite{conf/uss/ClementsGSGFKVB20}  & $\beta$/$\phi$ & $\varepsilon$/$\phi$ & $\varepsilon$ & $\varepsilon$/$\phi$  & 2, 3 & ARM, MIPS & \ding{109} & \ding{55} & Fuzzing \\
  FirmWire~\cite{conf/ndss/hernandez22} & $\beta$/$\phi$ & $\varepsilon$/$\phi$ & $\varepsilon$/$\phi$ & $\varepsilon$/$\phi$  & 2, 3 & ARM, MIPS & \ding{109} & \ding{55} & Fuzzing \\
  Jetset~\cite{conf/uss/JohnsonBZMCSL21} & $\phi$/$\chi$ & $\varepsilon$ & $\varepsilon$ & $\varepsilon$ & 2, 3 & ARM, Coldfire, x86 & \ding{119} & \ding{55} & Fuzzing \\
  Laelaps~\cite{conf/acsac/CaoGM020} & $\chi$ & $\varepsilon$ & $\varepsilon$ & $\varepsilon$ & 2, 3 & ARM & \ding{119} & \ding{55} & Fuzzing \\
  P$^2$IM~\cite{conf/uss/FengML20} & $\beta$/$\phi$ & $\varepsilon$ & $\varepsilon$ & $\varepsilon$ & 2, 3 & ARM & \ding{119} & \ding{55} & Fuzzing \\
  $\mu$Emu~\cite{conf/uss/Zhou21} & $\chi$ & $\varepsilon$ & $\varepsilon$ & $\varepsilon$ & 2, 3 & ARM & \ding{119} & \ding{55} & Fuzzing \\
  Ghidra EmulatorHelper & $\beta$ & $\varepsilon$ & $\varepsilon$ & $\varepsilon$ & 3 & \ding{92} & -  & \ding{55} & Generic\\
  \midrule
  \toolname                    & $\beta$/$\phi$/$\chi$ & $\varepsilon$/$\phi$ & $\varepsilon$ & $\varepsilon$/$\phi$   & 2,3 & \ding{92}                 & \ding{119}/\ding{109}       & \ding{51} & Generic (+inter-device) \\
  \bottomrule
\end{tabular}
\\{
  \vspace{4pt}
  \centering
  $\ddagger$: \textsc{Avatar} and \textsc{Avatar}$^2$ can, in principle, facilitate multi-device analysis by instantiating separate `System`/`Avatar` objects for each device.\\
  \ding{92}: \toolname supports all architectures supported by Ghidra, including: ARM, AArch64, C166, MIPS, PPC, SH-2, RH850, V850, x86, x86-64.\\
}
\vspace{-1em}
\end{table*}
\section{Source-code and data-sets}
\label{sec:open-source}

\begin{center}
\footnotesize
\begin{tabular}{p{7cm} | p{10cm}}
\toprule
Artifact & URL \\
\midrule
Core components & \url{https://anonymous.4open.science/r/metaemu-A670} \\
Static analysis component &  \url{https://anonymous.4open.science/r/metaemu-25CA} \\
Executor \& hooking components &   \url{https://anonymous.4open.science/r/metaemu-CC72} \\
Concolic executor components & \url{https://anonymous.4open.science/r/metaemu-3767} \\
Peripheral handling components & \url{https://anonymous.4open.science/r/metaemu-3F05} \\
Utility components (incl. peripherals and observers) & \url{https://anonymous.4open.science/r/metaemu-00B7}\\
\midrule
Firmware databases & \url{https://mega.nz/folder/JotCnSQK#bwCZjBjFmwnMqbTyhYWejQ}\\
\midrule
Open-source release & \url{https://fugue.re/}\\
\bottomrule
\end{tabular}
\end{center}

\end{document}